\newcount\mgnf  
\newcount\griglia
\mgnf=0

\ifnum\mgnf=0
\def\openone{\leavevmode\hbox{\ninerm 1\kern-3.3pt\tenrm1}}%
\def\*{\vglue0.3truecm}\fi
\ifnum\mgnf=1
\def\openone{\leavevmode\hbox{\ninerm 1\kern-3.63pt\tenrm1}}%
\def\*{\vglue0.5truecm}\fi

\ifnum\mgnf=0
   \magnification=\magstep0
   \hsize=14truecm\vsize=24.truecm
   \parindent=0.3cm\baselineskip=0.45cm
\font\titolo=cmbx12
\font\titolone=cmbx10 scaled\magstep 2
\font\cs=cmcsc10

\font\ottorm=cmr8

\font\msytw=msbm10

\font\indbf=cmbx10 scaled\magstep1

\fi
\ifnum\mgnf=1
   \magnification=\magstep1\hoffset=0.truecm
   \hsize=14truecm\vsize=24.truecm
   \baselineskip=18truept plus0.1pt minus0.1pt \parindent=0.9truecm
   \lineskip=0.5truecm\lineskiplimit=0.1pt      \parskip=0.1pt plus1pt
\font\titolo=cmbx12 scaled\magstep 1
\font\titolone=cmbx10 scaled\magstep 3
\font\cs=cmcsc10 scaled\magstep 1
\font\ottorm=cmr8 scaled\magstep 1

\font\msytw=msbm10 scaled\magstep1

\font\indbf=cmbx10 scaled\magstep2
\fi

\global\newcount\numsec\global\newcount\numapp
\global\newcount\numfor\global\newcount\numfig\global\newcount\numsub
\global\newcount\numlemma\global\newcount\numtheorem\global\newcount\numdef
\global\newcount\appflag
\numsec=0\numapp=0\numfig=1
\def\veroparagrafo{\number\numsec}\def\veraformula{\number\numfor}
\def\veraappendice{\number\numapp}\def\verasub{\number\numsub}
\def\verafigura{\number\numfig}
\def\verolemma{\number\numlemma}
\def\verotheorem{\number\numtheorem}
\def\veradef{\number\numdef}

\def\section(#1,#2){\advance\numsec by 1\numfor=1\numsub=1%
\numlemma=1\numtheorem=1\numdef=1\appflag=0%
\SIA p,#1,{\veroparagrafo} %
\write15{\string\Fp (#1){\secc(#1)}}%
\write16{ sec. #1 ==> \secc(#1)  }%
\hbox to \hsize{\titolo\hfill \number\numsec. #2\hfill%
\expandafter{\alato(sec. #1)}}\*}

\def\appendix(#1,#2){\advance\numapp by 1\numfor=1\numsub=1%
\numlemma=1\numtheorem=1\numdef=1\appflag=1%
\SIA p,#1,{A\veraappendice} %
\write15{\string\Fp (#1){\secc(#1)}}%
\write16{ app. #1 ==> \secc(#1)  }%
\hbox to \hsize{\titolo\hfill Appendix A\number\numapp. #2\hfill%
\expandafter{\alato(app. #1)}}\*}

\def\senondefinito#1{\expandafter\ifx\csname#1\endcsname\relax}

\def\SIA #1,#2,#3 {\senondefinito{#1#2}%
\expandafter\xdef\csname #1#2\endcsname{#3}\else
\write16{???? ma #1#2 e' gia' stato definito !!!!} \fi}

\def \Fe(#1)#2{\SIA fe,#1,#2 }
\def \Fp(#1)#2{\SIA fp,#1,#2 }
\def \Fg(#1)#2{\SIA fg,#1,#2 }
\def \Fl(#1)#2{\SIA fl,#1,#2 }
\def \Ft(#1)#2{\SIA ft,#1,#2 }
\def \Fd(#1)#2{\SIA fd,#1,#2 }

\def\etichetta(#1){%
\ifnum\appflag=0(\veroparagrafo.\veraformula)%
\SIA e,#1,(\veroparagrafo.\veraformula) \fi%
\ifnum\appflag=1(A\veraappendice.\veraformula)%
\SIA e,#1,(A\veraappendice.\veraformula) \fi%
\global\advance\numfor by 1%
\write15{\string\Fe (#1){\equ(#1)}}%
\write16{ EQ #1 ==> \equ(#1)  }}

\def\getichetta(#1){
\SIA g,#1,{\verafigura} %
\global\advance\numfig by 1%
\write15{\string\Fg (#1){\graf(#1)}}%
\write16{ Fig. #1 ==> \graf(#1) }}

\def\etichettap(#1){%
\ifnum\appflag=0{\veroparagrafo.\verasub}%
\SIA p,#1,{\veroparagrafo.\verasub} \fi%
\ifnum\appflag=1{A\veraappendice.\verasub}%
\SIA p,#1,{A\veraappendice.\verasub} \fi%
\global\advance\numsub by 1%
\write15{\string\Fp (#1){\secc(#1)}}%
\write16{ par #1 ==> \secc(#1)  }}

\def\etichettal(#1){%
\ifnum\appflag=0{\veroparagrafo.\verolemma}%
\SIA l,#1,{\veroparagrafo.\verolemma} \fi%
\ifnum\appflag=1{A\veraappendice.\verolemma}%
\SIA l,#1,{A\veraappendice.\verolemma} \fi%
\global\advance\numlemma by 1%
\write15{\string\Fl (#1){\lm(#1)}}%
\write16{ lemma #1 ==> \lm(#1)  }}

\def\etichettat(#1){%
\ifnum\appflag=0{\veroparagrafo.\verotheorem}%
\SIA t,#1,{\veroparagrafo.\verotheorem} \fi%
\ifnum\appflag=1{A\veraappendice.\verotheorem}%
\SIA t,#1,{A\veraappendice.\verotheorem} \fi%
\global\advance\numtheorem by 1%
\write15{\string\Ft (#1){\thm(#1)}}%
\write16{ th. #1 ==> \thm(#1)  }}

\def\etichettad(#1){%
\inum\appflag=0{\veroparagrafo.\veradef}%
\SIA d,#1,{\veroparagrafo.\veradef} \fi%
\inum\appflag=1{A\veraappendice.\veradef}%
\SIA d,#1,{A\veraappendice.\veradef} \fi%
\global\advance\numdef by 1%
\write15{\string\Fd (#1){\defz(#1)}}%
\write16{ def. #1 ==> \defz(#1)  }}

\def\Eq(#1){\eqno{\etichetta(#1)\alato(#1)}}
\def\eq(#1){\etichetta(#1)\alato(#1)}
\def\eqg(#1){\getichetta(#1)\alato(fig #1)}
\def\sub(#1){\0\palato(p. #1){\bf \etichettap(#1)\hskip.3truecm}}
\def\lemma(#1){\0\palato(lm #1){\cs Lemma \etichettal(#1)\hskip.3truecm}}
\def\theorem(#1){\0\palato(th #1){\cs Theorem \etichettat(#1)%
\hskip.3truecm}}
\def\definition(#1){\0\palato(df #1){\cs Definition \etichettad(#1)%
\hskip.3truecm}}

\def\equv(#1){\senondefinito{fe#1}$\clubsuit$#1%
\write16{eq. #1 non e' (ancora) definita}%
\else\csname fe#1\endcsname\fi}
\def\grafv(#1){\senondefinito{fg#1}$\clubsuit$#1%
\write16{fig. #1 non e' (ancora) definito}%
\else\csname fg#1\endcsname\fi}

\def\secv(#1){\senondefinito{fp#1}$\clubsuit$#1%
\write16{par. #1 non e' (ancora) definito}%
\else\csname fp#1\endcsname\fi}

\def\lmv(#1){\senondefinito{fl#1}$\clubsuit$#1%
\write16{lemma #1 non e' (ancora) definito}%
\else\csname fl#1\endcsname\fi}

\def\thmv(#1){\senondefinito{ft#1}$\clubsuit$#1%
\write16{th. #1 non e' (ancora) definito}%
\else\csname ft#1\endcsname\fi}

\def\defzv(#1){\senondefinito{fd#1}$\clubsuit$#1%
\write16{def. #1 non e' (ancora) definito}%
\else\csname fd#1\endcsname\fi}

\def\equ(#1){\senondefinito{e#1}\equv(#1)\else\csname e#1\endcsname\fi}
\def\graf(#1){\senondefinito{g#1}\grafv(#1)\else\csname g#1\endcsname\fi}
\def\secc(#1){\senondefinito{p#1}\secv(#1)\else\csname p#1\endcsname\fi}
\def\lm(#1){\senondefinito{l#1}\lmv(#1)\else\csname l#1\endcsname\fi}
\def\thm(#1){\senondefinito{t#1}\thmv(#1)\else\csname t#1\endcsname\fi}
\def\defz(#1){\senondefinito{d#1}\defzv(#1)\else\csname d#1\endcsname\fi}
\def\sec(#1){{\S\secc(#1)}}

\def\BOZZA{
\def\alato(##1){\rlap{\kern-\hsize\kern-1.2truecm{$\scriptstyle##1$}}}
\def\palato(##1){\rlap{\kern-1.2truecm{$\scriptstyle##1$}}}
}

\def\alato(#1){}
\def\galato(#1){}
\def\palato(#1){}

{\count255=\time\divide\count255 by 60 \xdef\hourmin{\number\count255}
        \multiply\count255 by-60\advance\count255 by\time
   \xdef\hourmin{\hourmin:\ifnum\count255<10 0\fi\the\count255}}

\def\oramin{\hourmin }

\def\data{\number\day/\ifcase\month\or gennaio \or febbraio \or marzo \or
aprile \or maggio \or giugno \or luglio \or agosto \or settembre
\or ottobre \or novembre \or dicembre \fi/\number\year;\ \oramin}
\setbox200\hbox{$\scriptscriptstyle \data $}
\footline={\rlap{\hbox{\copy200}}\tenrm\hss \number\pageno\hss}

\let\a=\alpha \let\b=\beta  \let\g=\gamma     \let\d=\delta  \let\e=\varepsilon
  \let\h=\eta    \def\th{\theta}
   \let\l=\lambda
        \let\x=\xi        \let\p=\pi      \let\r=\rho
\let\s=\sigma \let\t=\tau        
   \let\o=\omega 
      \let\L=\Lambda  
           
\let\O=\Omega 

\def\\{\hfill\break} \let\==\equiv

\let\io=\infty 

\let\0=\noindent

\def\ie{\hbox{\it i.e.\ }}
\def\der{\hbox{\rm d}}
\let\dpr=\partial 
\let\bs=\backslash

\def\tende#1{\,\vtop{\ialign{##\crcr\rightarrowfill\crcr
 \noalign{\kern-1pt\nointerlineskip}
 \hskip3.pt${\scriptstyle #1}$\hskip3.pt\crcr}}\,}
\def\otto{\,{\kern-1.truept\leftarrow\kern-5.truept\to\kern-1.truept}\,}

\def\PP{{\cal P}}\def\EE{{\cal E}}\def\VV{{\cal V}}

\def\TT{{\cal T}}
\def\RR{{\cal R}}\def\LL{{\cal L}}
\def\DD{{\cal D}}

 \def\acapo{\hfill\break}
  
\def\indica{\leaders \hbox to 0.5cm{\hss.\hss}\hfill}
\def\guida{\leaders\hbox to 1em{\hss.\hss}\hfill}
\mathchardef\oo= "0521

\def\xx{{\bf x}}
\def\yy{{\bf y}}\def\kk{{\bf k}}
\def\uu{{\bf u}}
 \def\bP{{\bf P}}\def\rr{{\bf r}}
\def\tt{{\bf t}}

\def\Halmos{\hfill\vrule height6pt width4pt depth2pt \par\hbox to \hsize{}}
\def\virg{\quad,\quad}
\def\proof{\0{\cs Proof - } }

\def\oo{{\underline \omega}}

\def\qed{\raise1pt\hbox{\vrule height5pt width5pt depth0pt}}

\def\indic{\hbox{\raise-2pt \hbox{\indbf 1}}}

\def\RRR{\hbox{\msytw R}}

 \def\ZZZ{\hbox{\msytw Z}}

%
%
%
\def\ins#1#2#3{\vbox to0pt{\kern-#2 \hbox{\kern#1 #3}\vss}\nointerlineskip}
%
%
%
\newdimen\xshift \newdimen\xwidth \newdimen\yshift
 
\def\insertplot#1#2#3#4#5{\par%
\xwidth=#1 \xshift=\hsize \advance\xshift by-\xwidth \divide\xshift by 2%
\yshift=#2 \divide\yshift by 2%
\line{\hskip\xshift \vbox to #2{\vfil%
#3 \includegraphics{#4.ps}}\hfill \raise\yshift\hbox{#5}}}
 
\def\initfig#1{%
\catcode`\%=12\catcode`\{=12\catcode`\}=12
\catcode`\<=1\catcode`\>=2
\openout13=#1.ps}
 
\def\endfig{%
\closeout13
\catcode`\%=14\catcode`\{=1
\catcode`\}=2\catcode`\<=12\catcode`\>=12}
 
 
\initfig{fig51}
\write13<
\write13<
\write13<gsave .5 setlinewidth 40 20 260 {dup 0 moveto 140 lineto} for stroke
grestore>
\write13</punto { gsave  
\write13<2 0 360 newpath arc fill stroke grestore} def>
\write13<40 75 punto>
\write13<60 75 punto>
\write13<80 75 punto>
\write13<100 75 punto 120 68 punto 140 61 punto 160 54 punto 180 47 punto 200
40 punto>
\write13<220 33 punto 240 26 punto 260 19 punto>
\write13<120 82.5 punto>
\write13<140 90 punto>
\write13<160 80 punto>
\write13<160 100 punto>
\write13<180 110 punto>
\write13<180 70 punto>
\write13<200 60 punto>
\write13<200 120 punto>
\write13<220 110 punto>
\write13<220 50 punto>
\write13<240 100 punto>
\write13<240 60 punto>
\write13<120 50 punto>
\write13<260 20 punto>
\write13<240 40 punto>
\write13<240 50 punto>
\write13<260 70 punto>
\write13<200 80 punto>
\write13<260 90 punto>
\write13<260 110 punto>
\write13<220 130 punto>
\write13<40 75 moveto 100 75 lineto 140 90 lineto 200 120 lineto 220 130
lineto>
\write13<200 120 moveto 240 100 lineto 260 110 lineto>
\write13<240 100 moveto 260 90 lineto>
\write13<140 90 moveto 180 70 lineto 200 80 lineto>
\write13<180 70 moveto 220 50 lineto 260 70 lineto>
\write13<220 50 moveto 240 40 lineto>
\write13<220 50 moveto 240 50 lineto>
\write13<100 75 moveto 260 20 lineto>
\write13<100 75 moveto 120 50 lineto stroke>
\write13<grestore>
\endfig

\openin14=\jobname.aux \ifeof14 \relax \else
\input \jobname.aux \closein14 \fi
\openout15=\jobname.aux
{\baselineskip=24pt
\centerline{\titolone Marginal Fermi liquid behaviour in the}
\centerline{\titolone $d=2$ Hubbard model with cut-off}
\vskip1.truecm
\centerline{\titolo{V. Mastropietro}}
\centerline{Dipartimento di Matematica, Universit\`a di Roma ``Tor 
Vergata''}
\centerline{Via della Ricerca Scientifica, I-00133, Roma}
\vskip1.truecm
\line{\vtop{
\line{\hskip1.5truecm\vbox{\advance \hsize by -3.1 truecm
\0{\cs Abstract 1.}
{\it We consider the half-filled Hubbard model with
a cut-off forbidding momenta close to the angles of the square
shaped Fermi surface. 
By Renormalization Group methods we find a convergent expansion
for the Schwinger function up to exponentially small
temperatures. We prove that
the system is not a Fermi liquid,
but on the contrary it behaves like a Marginal Fermi liquid,
a behaviour observed in the normal phase
of high $T_c$ superconductors.
}}
\hfill} }}
}
\vskip.5cm
\section(1,Main results)
\vskip.5cm
\*
\sub(1.1) {\it Motivations.} 
The notion of {\it Fermi liquids}, introduced by Landau,
refers to a wide class of interacting fermionic systems
whose thermodynamic properties (like the specific heat
or the resistivity)
are qualitatively the same 
of a gas of non interacting fermions.
While there is an enormous number of metals
having Fermi liquid behaviour, in recent times
new materials has been found whose properties are qualitatively
different. In particular
the high-temperature superconducting materials (so anisotropic
to be considered essentially bidimensional)
in their normal phase  have  a {\it non}
Fermi liquid behaviour, in 
striking contrast with 
previously known superconductors, 
which are Fermi liquids above the critical temperature.
While in Fermi liquids 
the {\it wave function renormalization} $Z$ 
is $Z=1+O(\l^2)$,
where $\l$ is the strength of the interaction,
in such metals it was found $Z\simeq 1+O(\l^2\log  T)$
for temperatures $T$ above the critical temperature, see [VLSAR]
(see also [VNS] for a review); metals behaving in this way were
called {\it Marginal Fermi liquids}.
Such results stimulated an intense theoretical research.
It was found by a perturbative analysis, 
see for instance [AGD] or [Sh], that
in a system of
weakly interacting fermions in $d=2$ 
$Z$ 
is essentially temperature independent,
at least for circular or "almost" circular Fermi surfaces.
Despite doubts appeared 
about the reliability of results obtained by
perturbative expansions [A], 
such results were indeed confirmed recently by 
rigorous Renormalization group methods.
It was proved in [FMRT] and [DR] 
that indeed a weakly interacting Fermi system
with a circular Fermi surface is a Fermi liquid, up to exponentially
small temperatures.    
Such result was extended in [BGM] to all possible
weakly interacting $d=2$ fermionic systems with symmetric, smooth and convex 
Fermi surfaces, up to exponentially small temperatures.
These results cannot be obtained by
dimensional power counting arguments as such arguments
give 
a bound $|Z-1|\le C\l^2|\log T|$ from which one cannot
distinguish Fermi or non Fermi liquid behaviour;
for obtaining $Z=1+O(\l^2)$
one has instead to use delicate
volume improvements
in the integrals expressing $Z$,
based on the geometrical constraints
to which the momenta close to the Fermi surface (assumed convex, regular and symmetric)
are subjected. 

As Fermi liquid behaviour is found in 
systems with symmetric, smooth and convex 
Fermi surfaces, in order to find non Fermi liquid behaviour 
one has to relax some of
such conditions.
It was pointed out, see for instance [VR] and [ZYD], that 
the presence in the Fermi surface 
of {\it flat} regions in opposite sides
could produce a non
Fermi liquid behaviour; flat regions are indeed present
in the Fermi surfaces
of high $T_c$ superconductors [S]. 
The simplest model exhibiting a Fermi surface
with flat pieces is the {\it half-filled Hubbard model},
describing a system of spinning $d=2$ fermions with local
interaction and dispersion relation given by
$\e(k_x,k_y)=\cos k_x+\cos k_y$. The Fermi surface
is the set of momenta such that 
$\e(k_x,k_y)=0$ and it is a square with corners $(\pm\pi,0)$
and $(0,\pm\pi)$. However this model
has the complicating feature of
vanishing Fermi velocity at the points $(\pm\pi,0)$
and $(0,\pm\pi)$ \ie at the corners of the Fermi surface;
this originates to the so called {\it Van Hove singularities}
in the density of states. 
In order to investigate the
possible non Fermi liquid behaviour
of interacting fermions with a Fermi surface with flat
pieces, independently
from the presence of Van Hove singularities,
one can introduce in the half filled
Hubbard model a cut-off forbidding
momenta near the corners of the Fermi surface.
The half filled
Hubbard model with cut-off 
(or the essentially equivalent, but slightly simpler, 
problem of fermions
with the linearized dispersion relation
$\e(k_x,k_y)=|k_x|+|k_y|-\pi$)
has been extensively studied in literature,
see for instance
[M], [L], [ZYD],[VR],[FSW], [DAD], [FSL].
The cut-off is somewhat artificially introduced
but the idea is that the model, at least for same values of the
parameters, belongs to
the same university 
class of models with "almost" squared and smooth Fermi surface, like
the anisotropic Hubbard
models [Sh], the
Hubbard model with nearest and next to
nearest neighbor
interaction [M], or the half-filled Hubbard model
close to half filling.

Aim of this paper is to
compute in a rigorous way
the asymptotic behaviour of the Schwinger functions
of the half filled Hubbard model with cut-off
up to exponentially small temperatures.
We will show that such a system is indeed a Marginal
Fermi liquid, and our result furnishes indeed the first example
rigorously established of such behaviour in $d=2$. 

For our convenience, we will consider new variables 
$k_+={k_x+k_y\over 2}$ and $k_-={k_x-k_y\over 2}$ so that
the dispersion relation of the half-filled Hubbard model
is given by
$$\e(k_+,k_-)=2\cos k_+\cos k_-\Eq(qed)$$ 
and the Fermi surface is the set
$k_+=\pm{\pi\over 2}$ or $k_-=\pm{\pi\over 2}$.

\*\sub(1.2) {\it The model.}
Given a square
$[0,L]^2\in\RRR^2$, the inverse temperature $\b$ and the (large) integer $M$,
we introduce in $\L=[0,L]^2\times[0,\b]$ a lattice $\L_M$, whose sites are
given by the 
{\it space-time points} $\xx=(x_0,x_+,x_-)$ with $(x_{+},x_-)\in \ZZZ^2$
and $x_0=
n_0
\b/M$, $n_0=0,1,\ldots,M-1$. We also consider the
set $\DD$ of {\it space-time momenta} 
$\kk=(k_0,k_+,k_-)\equiv(k_0,\vec k)$, with 
$k_{\pm},={2\p n_{\pm} \over L}$, $(n_+,n_-)\in\ZZZ^2$, 
$[-L/2]\le n_{\pm}\le[L-1/2]$;
$k_0={2\pi\over \b}(n_0+{1\over 2})$,
$n_0=0,1,\ldots,M-1$.
With each $\kk\in\DD$ we associate four
Grassmanian variables $\hat\psi^\e_{\kk, s}$, $\e, s\in\{+,-\}$;
$s$ is the spin. The
lattice $\L_M$ is introduced only for technical reasons so that the number of
Grassmann variables is finite, and eventually the (essentially trivial) limit
$M\to\io$ is taken.
We introduce also a linear functional $P(d\psi)$ on
the Grassmanian algebra generated by the variables $\hat\psi^\e_{\kk,\s}$,
such that
$$\int P(d \psi) \hat \psi^-_{\kk_1, s_1}\hat \psi^+_{\kk_2, s_2} = L^2\b
\d_{s_1, s_2}\d_{\kk_1,\kk_2}
\hat g(\kk_1)\;,\Eq(2.1)$$
where $g(\kk)$ is defined by 
$$\hat g(\kk)={\chi(\kk)\over
-i k_0+2\cos k_+\cos k_-}\Eq(110)$$
where $\chi(\kk)$ is a cut-off function 
$$\chi(\kk)=H(a_0^2\sin^2 k_+)
C_0^{-1}(\kk)+H(a_0^2\sin^2 k_-)
C_0^{-1}(\kk)\Eq(101)$$
where
$$C_0^{-1}(\kk)=H(\sqrt{k_0^2+4\cos^2(k_+)\cos^2(k_-)}
)\Eq(102)$$
and, if $\g> 1$ and $a_0\ge \sqrt{2}$
$$ H(t) = \cases{
1 & if $|t| <\g^{-1} \;$ \cr
0 & if $|t| >1\; $\cr},\Eq(2.30)$$
The function $C_0^{-1}(\kk)$ acts as an ultraviolet cut-off
forcing the momenta $\vec k$ to be not too far from the Fermi surface,
and $k_0$ not too large; the cut-off on $k_0$ is imposed
only for technical convenience and it could be easily removed.
The functions $H(a_0^2\sin^2 k_{\pm})$ forbids momenta 
near the corners of the Fermi surface \ie the 
points $(\pm\pi/2,\pm\pi/2)$. 
The {\sl Grassmanian field} $ \psi^\e_\xx$ is defined by
$$\psi_{\xx, s}^{\pm}= {1\over L^2\b}
\sum_{\kk\in {\cal D}}\hat \psi_{\kk,s}^{\pm}
e^{\pm i\kk\cdot\xx}\; .\Eq(2.4)$$

The ``Gaussian measure'' $P(d \psi)$ has a simple representation in terms of
the ``Lebesgue Grassmanian measure''
$$\hbox{\sl D}\psi=\prod^*_{\kk\in\DD, s=\pm}
d\hat\psi_{\kk,s}^+ d\hat\psi_{\kk,s}^-,\Eq(2.5)$$
defined as the linear functional on the Grassmanian algebra, such that, given
a monomial $Q( \hat\psi^-, \hat\psi^+)$ in the variables $\hat\psi_{\kk,s}^-,
\hat\psi_{\kk,s}^+$, its value is $0$, except in the case 
$Q( \hat\psi^-,
\hat\psi^+)= \prod^*_{\kk,s} \hat\psi^-_{\kk,s} \hat\psi^+_{\kk,s}$, up to a
permutation of the variable, in which case its value is $1$. 
Finally 
$\prod^*_{\kk\in\DD, s=\pm}$
means a product over the $\kk$ such that
$\chi(\kk)>0$.
We define
$$
P(d\psi) = N^{-1} \hbox{\sl D}\psi \cdot\;\exp [-{1\over L^2\b}
\sum^*_{\kk\in\DD,\s=\pm } 
\chi^{-1}(\kk)(-i k_0+2\cos k_+\cos k_-)
\hat\psi^{+}_{\kk,s}\hat\psi^{-}_{\kk,s}]\;,\Eq(2.6)$$
with $N$ is a renormalization constant and again 
$\sum^*_\kk$ means a sum over $\kk$ such that $\chi(\kk)>0$.
 
The two point {\it Schwinger function} is defined by the following
{\it Grassman functional integral}
$$S(\xx-\yy)=\lim_{L\to\io}\lim_{M\to\io}{
\int P(d\psi)
e^{-\VV(\psi)}\psi^{-}_{\xx,s}\psi^+_{\yy,s}
\over \int P(d\psi) e^{-\VV(\psi)}}\;,\Eq(2.7)$$
where, if we use $\int d\xx$ as a shorthand for 
${\b\over M}\sum_{\xx\in \Lambda_M}
$,
$$\VV(\psi)=\l\sum_{s}\int d\xx \psi^+_{\xx,s}
\psi^-_{\xx,s} \psi^+_{\xx,-s}\psi^-_{\xx,-s}\;,\Eq(2.8)$$
We call $\hat S(\kk)$ the Fourier transform of $S(\xx-\yy)$.
\*\sub(1.3) {\it Main Theorem.}
Our main results are summarized by the following Theorem, 
which will be proved in the following sections.
\vskip.5cm
{\bf Theorem.} {\it  Given $a_0$ large enough,
there exist two positive constants $\e$ and $\bar c$
such that, for all $|\l|\le\e$ and $T\ge \exp\{-
(\bar c |\l|)^{-1}\}$, for all $\kk\in\DD$ such that
${\pi\over2\b}\le\sqrt{k_0^2+4\cos^2 k_+\cos^2 k_-}
\le {3\pi\over 2\b}$ 
and $H(a_0^2\sin^2 k_-)=1$ then
$$\hat S(\kk)=
{(k_0^2+4\cos^2 k_+ \cos^2 k_-)^{\h(k_-)}
\over -i k_0 
+2\cos k_+ \cos k_-}(1+\l^2 A_{I}(\kk))\Eq(sf1)$$
and 
for $\kk\in\DD$ such that
${\pi\over2\b}\le\sqrt{k_0^2+4\cos^2 k_+\cos^2 k_-)}
\le {3\pi\over 2\b}$ 
and $H(a_0^2 \sin^2 k_+)=1$ then
$$\hat S(\kk)=
{(k_0^2+4\cos^2 k_+ \cos^2 k_-)^{\h(k_+)}
\over -i  k_0 
+2\cos k_+ \cos k_-}(1+\l^2 A_{II}(\kk))\Eq(sf2)$$
where
$|A_{i}(\kk)|\le c$, where $c>0$ is a constant,
 and $\h(k_\pm)=a(k_\pm)\l^2+O(\l^3)$ is a critical index expressed 
by a convergent series with $a(k_\pm)\ge 0$ 
a not identically vanishing smooth function.}
\*\sub(1.5) {\it Remarks.}
The above theorem describes the behaviour of the
two point Schwinger function up to exponentially small temperatures, \ie 
$T\ge \exp\{-
(\bar c |\l|)^{-1}\}$; the constant $\bar c$ is essentially
given by the second order terms of the perturbative expansion.
A straightforward consequence of \equ(sf1), \equ(sf2) is that
that the wave function renormalization is
$Z=1+O(\l^2\log\b)$, which means that 
the half-filled Hubbard model
with cut-off is a 
marginal Fermi liquid up to exponentially small temperatures. 
From \equ(sf1), \equ(sf2) we see that
the behaviour of the Schwinger function close to
the Fermi surface is {\it anomalous} and described
by {\it critical indices} which are functions of the projection
of the momentum on the Fermi surface.
Critical indices which are momentum dependent were found
for the same model
also in [FSL] by heuristic bosonization methods.
The presence of the critical indices
makes the Schwinger function quite similar to the one for $d=1$
interacting spinless fermionic systems, characterized by {\it
Luttinger liquid} behaviour (see for instance [A]). 
However an important difference is that
the critical exponent $\h$ in a Luttinger liquid
is a number, while here is a function of the momenta.
Another crucial difference is that
in a 
{\it Luttinger liquid}
$\hat S(\kk)\simeq \hat g(\kk)|\kk|^\h$, with $\h=a\l^2+O(\l^3)$
{\it up to $T=0$};
hence a Luttinger liquid is a Marginal Fermi liquid
for high enough temperatures but not all the marginal
Fermi liquids are Luttinger liquids. 

The paper is organized in the following way.
In \S 2 we implement Renormalization Group ideas
by writing the Grassman integration in (1.10)
as the product of many integrations at
different scales. The integration 
of a single scale leads to new effective interactions, 
and the {\it renormalization} consists in subtracting
from the kernels of the effective interaction, which are not dimensionally
irrelevant, of the effective interaction their value
computed at the Fermi surface. 
One obtains
an expansion for the
Schwinger functions as {\it power series}
of a set of {\it running couplings functions} (depending from the momentum
on the Fermi surface and the scale). In \S 3 we prove that
this series is convergent if the running coupling
functions are small enough; the convergence radius is
finite and temperature independent, and this  
means that the theory is {\it renormalizable}. 
In the proof of convergence 
one uses the Gram-Hadamard inequality.
In \S 4 we show that the running coupling functions obey to a 
recursive set of integral equation, called
{\it Beta function}, and we show that
the running coupling functions remain small
up to exponentially small temperatures $T\ge \exp\{-
(\bar c |\l|)^{-1}\}$. Moreover we show that the 
wave function renormalization has an anomalous flow,
with a non vanishing exponent (contrary to what
happens for instance in the case of circular Fermi surfaces),
and this essentially concludes the proof of the Theorem.
It would be possible to use our beta
function 
to detect (at least numerically)
the main instabilities of the system at very low temperatures.
At the moment, 
this kind of numerical analysis
was done for this model 
only in [ZYD] in the {\it parquet approximations}, with
no control on higher orders which are simply neglected.
Finally in \S 5 we compare the Marginal Fermi liquid behaviour
we find in this model with the Luttinger liquid behaviour, and
we discuss briefly what happens in the Hubbard model
with cut-off {\it close} to half filling.

It is very likely that the half-filled Hubbard model 
with cut-off can work
as a paradigm for a large class of systems, in which
the Fermi surface is flat or almost flat but there are
no Van Hove singularities. Marginal Fermi liquid
behaviour can be surely found in the Hubbard
model with cut-off and close to half-filling,
up to temperatures above the inverse of the radius
of curvature of the Fermi surface. 
Another model in which one could possibly find
Marginal Fermi liquid behaviour is the
anisotropic
Hubbard model introduced in [Sh] with dispersion relation 
$\cos k_1+t \cos k_2$, with $t=1+\e$.
Such model has a Fermi surface with no van Hove singularities
and four "almost"
flat and parallel pieces, and one can expect
$Z=1+O(\l^2|\log(|\e|)|\log\b)$ for $\b\le O(\min[\e^{-1},
\exp\{
(\bar c |\l|)\}]$.
Another interesting question
is the possibility of Marginal Fermi liquid behaviour
in the Hubbard model {\it close} to half-filling (with no cut-off).
At half-filling it is believed
$Z\simeq 1+O(\l^2\log^2\b)$, so a different behaviour
with respect to
Marginal Fermi liquid behaviour.
A renormalization group analysis for this problem 
was begun in [R], and it was proved the
convergence of the series not containing
subgraphs with two external lines 
for $T\ge \exp\{-(c_0 |\l|)^{-1\over 2}\}$.

\vskip.5cm
\section(2, Renormalization Group analysis)
\vskip.5cm
\*
\sub(2.1) {\it The scale decomposition.}
As the spin index will play no role in the following analysis 
(on the contrary
it is expected to have an important role at lower temperatures) we simply omit
it. The cut-off function $\chi(\kk)$ defined in 
\equ(101) has a support in the $\vec k$ space
which is given by four disconnected regions, each one containing
only one flat side of the Fermi surface. It is natural then to write
each Grassman variable as a sum of four independent
Grassman variables, with momentum $\vec k$
having value in one of the four disconnected regions;
each field will be labeled by a couple of indices, $\s=I,II
$ and $\o=\pm 1$, so that each field has spatial momenta with values 
in the region containing $(\o p_F,0)$ if $\s=I$ or $(0,\o p_F)$ if $\s=II$.
We write then Grassman integration
as 
$$\int P(d\psi) F(\psi)=\int
\prod_{\s=I,II}\prod_{\o=\pm 1} P_{\s,\o}(d\psi)
F(\sum_{\s=I,II}\sum_{\o=\pm 1}\psi_{\s,\o})\Eq(neo44)$$
where $F$ is any monomial, $\o=\pm 1$ and 
$$\int P_{I,\o}(d\psi)
\hat\psi^-_{I,\o,\kk_1'+\o \vec p_{F,I}}
\hat\psi^+_{I,\o', \kk'+\o' \vec p_{F,I}}
=\d_{\o,\o'}\d_{\kk_1',\kk'}
H(a_0^2 \sin^2 k_-) {C_\o^{-1}(k_0,k'_+,k_-)
\over -i k_0+2\o  \sin k'_+ \cos k_-}\Eq(12)$$
$$\int P_{II,\o}(d\psi)
\hat\psi^-_{II,\o,\kk_1'+\o \vec p_{F,II}}
\hat\psi^+_{II,\o',\kk'+\o' \vec p_{F,II}}$$
$$=\d_{\kk_1',\kk'}\d_{\o,\o'}
H(a_0^2 \sin^2 k_+) {C_\o^{-1}(k_0,k'_-,k_+)
\over -i k_0+2\o  \sin k'_- \cos k_+}\Eq(13)$$
where 
$$C^{-1}_\o(k_0,k'_+,k_-)=
\th(\o k'_+ +p_F)H(\sqrt{k_0^2+4\sin^2 k'_+\cos k_-})\Eq(am1)$$
$$C^{-1}_\o(k_0,k'_-,k_+)=\th(\o k'_-+p_F)
H(\sqrt{k_0^2+4\sin^2 k'_-\cos k_+})\Eq(am2)$$
and $\vec p_{F,\s}$ is defined such that $\vec p_{F,I}=({\pi\over 2},0)$
and $\vec p_{F,II}=(0,{\pi\over 2})$;
moreover $p_F={\pi\over 2}$ and $\vec k=\vec k'+\o \vec p_{F,\s}$ (
$\vec k'$ is the momentum measured from the Fermi surface).

It is convenient,
for clarity reasons, to start by studying
the "free energy" of the model,
defined as
$$-{1\over L^2\b}\log\int P(d\psi)e^{-\VV}\Eq(cds)$$
where, calling with a slight abuse of notation
$\hat\psi_{\s,\o,\kk'+\o \vec p_{F,\s}}\equiv \hat\psi_{\s,\o,\kk'}$,
$\VV$ is equal to
$$\l\sum_{\o_1,..,\o_4}\sum_{\s_1,..,\s_4=I,II}\int d\kk'_1...d\kk'_4 
\d(\sum_{i=1}^4\e_i(\kk'_i+\o_i \vec p_{F,\s_i}))
\hat\psi^+_{\s_1,\o_1,\kk'_1}
\hat\psi^+_{\s_2,\o_2,\kk'_2}\hat\psi^-_{\s_3,\o_3,\kk'_3}
\hat\psi^-_{\s_4,\o_4,\kk'_4}\Eq(15)$$
where $\int d\kk={1\over L^2\b}\sum_\kk$ and
$\d(\kk-\kk')=L^2\b\d_{\kk,\kk'}$.

We will evaluate the Grassman integral \equ(cds) by a multiscale
analysis based on (Wilsonian) Renormalization Group ideas. 
The starting point is the following decomposition
of the cut-off functions \equ(am1), \equ(am2)
$$H(\sqrt{k_0^2+4\cos^2\hat k_\s\sin^2 \underline k'_\s})=\sum_{k=-\io}^0  \bar f_k(
\sqrt{k_0^2+4\cos^2\hat k_\s\sin^2 \underline k'_\s})
\equiv\sum_{k=-\io}^0 f_k(k_0,\underline k'_\s,\hat k_\s)
\Eq(dec)$$
with $\bar f_k(t)=H(\g^{-k}t)-H(\g^{-k+1}t)$
is a smooth compact support function,
with support $\g^{k-1}\le |t|\le\g^{k+1}$; moreover:

a)$\hat k_\s=k_-$ if $\s=I$ and
$\hat k_{\s}=k_+$ if $\s=II$; $\hat k_\s$ is
the projection of $\vec k$ in the direction {\it parallel}
to the Fermi surface.

b)$\underline k'_\s=k'_+$ if $\s=I$ and
$\underline k'_{\s}=k'_-$ if $\s=II$; 
$\underline k'_\s+\o p_{F,\s}$
is the projection of $\vec k$ in the direction {\it normal}
to the Fermi surface.

For each $\s$, 
the function $f_k(k_0,\underline k'_\s,\hat k_\s)$ has a support
in two regions of thickness $O(\g^h)$ 
around each flat side of the Fermi
surface, at a distance $O(\g^h)$ from it.
We will assume $L=\io$ for simplicity
and 
it follows that there is a $h_\b=O(\log\b)$
such that $f_k=0$ for $k< h_{\b}$, while
$f_k$ is not identically vanishing
for $k\ge h_\b$.

The integration of \equ(cds) will be done iteratively
integrating out the fields with momenta closer and closer
to the Fermi surface.
We will prove by induction that it is possible to
define a sequence of functions $Z_{h}(\bar k'_{\s,\o})$
and a sequence of {\it effective potentials}
$\VV^{(h)}$ such that
$$\int P_{I}(d\psi)P_{II}(d\psi)  
e^{-\VV}=e^{-L^2\b E_h}\int
P_{Z_h,I}(d\psi^{(\le h)})
P_{Z_h,II}(d\psi^{(\le h)})
e^{-\VV^{(h)}(\sqrt{Z_h}\psi^{(\le h)})}.\Eq(3.3a)$$
where $E_h$ is a constant and
$\sqrt{Z_h}\hat\psi^{(\le h)}$ equal to
$$(\sqrt{Z_h(\bar k'_{I,1})}\hat\psi_{I,1,\kk'}^{(\le h)},
\sqrt{Z_h(\bar k'_{I,-1})}
\hat\psi_{I,-1,\kk'}^{(\le h)},
\sqrt{Z_h(\bar k'_{II,1})}\hat\psi_{II,1,\kk'}^{(\le h)},
\sqrt{Z_h(\bar k'_{II,-1})}\hat\psi_{II,-1,\kk'}^{(\le h)})\Eq(16)$$
and
$P_{Z_h,\s}(d\psi^{(\le h)})$ is the fermionic integration
with propagator
$$g^{\le h}_{\s,\o}(\kk')=
{1\over Z_{h}(\bar k'_{\s,\o})}
 {H(a_0^2 \sin^2 \hat k_\s)
C_{h,\o}^{-1}(k_0,\underline k'_\s,\hat k_\s)\over 
-i k_0+2\o \cos \hat k_\s \sin \underline k'_\s}\th(\o k'_\s+\o p_F)\Eq(3.3)$$
with  
$$C_{h,\o}^{-1}(k_0,\underline k'_\s,\hat k_\s)=\sum_{k=-\io}^h 
f_k(k_0,\underline k'_\s,\hat k_\s). \Eq(3.4)$$
The $\th$-function in \equ(3.3) can be omitted
by the definition of the variables $\underline k'_\s$.

We define 
$\bar k_{\s,\o}=({\pi\over\b},\o p_F, k_{-})$ if $\s=I$ and 
$\bar k'_{\s,\o}=({\pi\over\b},k_{+},\o p_F)$ if $\s=II$;
moreover
$\bar k'_{\s,\o}=({\pi\over\b},0, k_{-})$ if $\s=I$ and 
$\bar k'_{\s,\o}=({\pi\over\b},k_{+},0)$ if $\s=II$; moreover
we call
$\bar k''_{\s,\o}=(-{\pi\over\b},0, k_{-})$ if $\s=I$ and 
$\bar k''_{\s,\o}=(-{\pi\over\b},k_{+},0)$ if $\s=II$.

If $\e=\pm$
$$\VV^{(h)}(\psi^{\le h})=\sum_{n=1}^\io\sum_{\o_1,...,\o_{2n}}
\sum_{\s_1,...,\s_{2n}}\sum_{\e_1,...,\e_n}$$
$$\int d\kk'_1...d\kk'_{2n} 
\d(\sum_{i}\e_i (\kk'_i+\o_i
\vec p_{F,\s_i}))
\left[\prod_{i=1}^{2n} 
\hat\psi^{(\le
h)\e_i}_{\s_i,\o_i,\kk'_i}\right]
\hat W^{(h)}_{2n}(\kk'_1,...,\kk'_{2n-1})\Eq(3.3aa)$$
where
$$\hat W^h_{2n}(\kk_1...\kk_{2n-1})=
\hat W^h_{2n}(\kk'_1+\o_1 \vec p_{F,\s_1}...\kk_{2n-1}+\o_{2n-1} \vec p_{F,\s_{2n-1}})
=\hat W^h_{2n}(\kk'_1...\kk'_{2n-1})
\Eq(91)$$
\*
\sub(2.2) {\it The renormalization procedure.}
Let us show that \equ(3.3a) is true for $h-1$, assuming that
it is true for $h$. 
We define an $\LL$ operator
acting linearly on the kernels of the effective potential \equ(3.3aa):
\vskip.5cm
1)$\LL \hat W^{(h)}_{2n}=0$ if $n\ge 2$
\vskip.5cm
2)If $n=1$ 
$$\LL\hat W^h_2(\kk')=
{1\over 2}[W^h_2(\bar k'_{\s,\o})+
\hat W^h_2(\bar k''_{\s,\o})]+k_0\partial_{k_0}
\hat W^h_2(\bar k'_{\s,\o})+
\sin \bar k'_\s \partial_{\s}
\hat W^h_2(\bar k'_{\s,\o})]\Eq(loc1a)$$
where $\partial_{k_0}$ means the discrete derivative
and $\partial_\s=\partial_{k_+}$ is $\s=I$ and 
$\partial_\s=\partial_{k_-}$ is $\s=II$.
We will prove in \S 4 that 
$[\hat W^h_2(\bar k'_{\s,\o})+
\hat W^h_2(\bar k''_{\s,\o})]=0$.
\vskip.5cm
3)If $n=2$
$$\LL
\hat W^h_4(\kk'_1,\kk'_2,\kk'_3)=
\hat W^h_4(\bar k'_{\s_1,\o_1},\bar k'_{\s_2,\o_2},
\bar k'_{\s_3,\o_3})
\Eq(loc2)
$$
\vskip.5cm
Calling 
$ \partial_{0}
\hat W^h_2(\bar k'_{\o,\s})=-i a_h(\bar k'_{\o,\s})$,  
$ \partial_{\s}
\hat W^h_2(\bar k'_{\o,\s})=2\o\cos \hat k_\s z_h(\bar k'_{\o,\s})$
and 
$$l_{h}(\bar k'_{\s_1,\o_1}, \bar k'_{\s_2,\o_2},
\bar k'_{\s_3,\o_3})=\hat W_4^h(\bar k'_{\s_1,\o_1}, \bar k'_{\s_2,\o_2},
\bar k'_{\s_3,\o_3})\Eq(into)$$
we can write
$$\LL \VV^h=\sum_{\s=I,II}\int d\kk'[
z_h(\bar k'_{\o,\s})2\o \cos \hat k_\s\sin \underline k'_\s-i k_0
a_h(\bar k'_{\o,\s})]\hat\psi^{+(\le h)}_{\kk',\s,\o}\hat\psi^{-(\le h)}_{\kk',\s,\o}+
\sum_{\{\o\},\{\s\} }$$
$$\int d\kk'_1...d\kk'_4
l_{h}(\bar k'_{\o_1,\s_1}, \bar k'_{\o_2,\s_2},
\bar k'_{\o_3,\s_3})
\hat\psi^{+(\le h)}_{\kk'_1,\s_1,\o_1}\hat\psi^{+(\le h)}_{\kk'_2,\s_2,\o_2}
\hat\psi^{-(\le h)}_{\kk'_3,\s_3,\o_3}\
\hat\psi^{-(\le h)}_{\kk'_4,\s_4,\o_4}\d(\sum_i\e_i(\kk'_i+p_{F,\s_i}))\Eq(loc3)$$

We write the r.h.s. of \equ(3.3a) as
$$\int
P_{I,Z_h}(d\psi^{(\le h)})\int
P_{II,Z_h}(d\psi^{(\le h)})
e^{-\LL\VV^{(h)}(\sqrt{Z_h}\psi^{(\le h)})-\RR\VV^{(h)}(\sqrt{Z_h}\psi^{(\le h)})}\Eq(amaba3)$$
with $\RR=1-\LL$.
\*
\sub(2.4) {\it Remark 1.} The non trivial action of $\RR$ 
on the kernel with $n=2$
can be written as
$$\RR \hat W^h_4(\kk'_1,\kk'_2,\kk'_3)=
[ \hat W^h_4(\kk'_1,\kk'_2,\kk'_3)- 
\hat W^h_4(\bar k'_{\s_1,\o_1},\kk'_2,\kk'_3)]$$
$$+[\hat W^h_4(\bar k'_{\s_1,\o_1},\kk'_2,\kk'_3)- 
\hat W^h_4(\bar k'_{\s_1,\o_1},\bar k'_{\s_2,\o_2},\kk'_3)]\Eq(21)$$
$$+
[ \hat W^h_4(\bar k'_{\s_1,\o_1},\bar k'_{\s_2,\o_2},\kk'_3)- 
\hat W^h_4(\bar k'_{\s_1,\o_1},
\bar k'_{\s_2,\o_2},\bar k'_{\s_3,\o_3})]$$
The first addend can be written as, if $\s_1=I$ (say), in the limit
$L\to\io$
$$(k_{0,1}-{\pi\over\b})\int_0^1 dt \partial_{k_{0,1}} 
\hat W_4^h({\pi\over\b}+t (k_{0,1}-{\pi\over\b}), k'_{+,1},k_{-,1};\kk'_2,\kk'_3)
+\Eq(22)$$
$$k'_{+,1}\int_0^1 dt \partial_{k'_{+,1}} 
\hat W_4^h(0, t k'_{+,1},k_{-,1};\kk'_2,\kk
'_3)$$
The factors $k_{0,1}-\pi/\b$ and $k'_{+,1}$ are $O(\g^{h'})$,
for the compact support properties of the propagator associated to
$\psi^{+(\le h)}_{I,\o_1,\kk'_1}$, with $h'\le h$,
while the derivatives are dimensionally $O(\g^{-h+1})$; hence the effect of $\RR$
is to produce  a factor $\g^{h'-h-1}<1$. Similar considerations
can be done for the other addenda and for the action of $\RR$ on the $n=1$ terms.

{\it Remark 2.} From \equ(loc2) we see that the effect of the 
$\LL$ operation is to 
replace in $W^h_2(\kk)$ the momentum $\vec k$ with its projection
on the closest flat side of the Fermi surface.
Hence the fact that  
the propagator is singular over an extended region
(the Fermi surface) and not simply in a point has the effect
that the renormalization point cannot be fixed but
it must be left moving on the Fermi surface.
\*
\sub(2.5) {\it The anomalous integration.}
In order to integrate the field $\psi^{(h)}$ 
we can write
$$\int
P_{I,Z_h}(d\psi^{(\le h)})\int
P_{II, Z_h}(d\psi^{(\le h)})
e^{-\LL\VV^{(h)}(\sqrt{Z_{h}}\psi^{(\le h)})
-\RR\VV^{(h)}(\sqrt{Z_{h}}\psi^{(\le h)})}=\Eq(am)$$
$$\int
P_{I,Z_{h-1}}(d\psi^{(\le h)})\int
P_{II,Z_{h-1}}(d\psi^{(\le h)})
e^{-\LL\tilde \VV^{h}(\sqrt{Z_{h}}\psi^{(\le h)})-
\RR\VV^{(h)}(\sqrt{Z_{h}}\psi^{(\le h)})}
$$
where $P_{\s,Z_{h-1}}(d\psi^{(\le h)})$ is the fermionic integration
with propagator
$${1\over Z_{h-1}(\kk')}
{H(a_0^2\sin^2 \hat k_\s)
C_h^{-1}(k_0,\underline k'_\s,\hat k_\s)
\over -i k_0+2\o \cos \hat k_{\s}\sin \underline k'_\s}\Eq(3.3kb)$$
and
$$Z_{h-1}(\kk')=Z_{h}(\bar k'_{\s,\o})[1+
H(a_0^2\sin^2 \hat k_\s)
C^{-1}_h(k_0,\underline k'_\s,\hat k_\s) a_h(\bar
k'_{\s,\o})]\Eq(33)$$
Moreover $$\LL\tilde \VV^{h}=\LL\VV^{h}-
\sum_{\s=I,II}\int d\kk'z_h(\bar k'_{\o,\s})[
2\o \cos \hat k_\s\sin \underline k'_\s-i k_0
]\hat\psi^{+(\le h)}_{\kk',\s,\o}\hat\psi^{-(\le h)}_{\kk',\s,\o}.\Eq(am2)$$

We rescale the fields by rewriting the r.h.s. of \equ(am)
as
$$\int
P_{I,Z_{h-1}}(d\psi^{(\le h)})\int
P_{II,Z_{h-1}}(d\psi^{(\le h)})
e^{-\LL\hat \VV^{h}(\sqrt{Z_{h-1}}\psi^{(\le h)})-
\RR\VV^{(h)}(\sqrt{Z_{h-1}}\psi^{(\le h)})}
\Eq(am1)$$
where
$$\LL \hat V^h=\int d\kk\sum_{\s=I,II}[
\d_{h,\o}(\bar k'_{\s})2 \o \cos\hat k_\s\sin \underline k'_\s)]
\hat\psi^+_{\s,\kk',\o}
\hat\psi^-_{\s,\kk',\o}+
\sum_{\s_1,..,\s_4=I,II}\int d\kk'_1..d\kk'_4$$
$$\l_{h}
(\bar k'_{\s_1,\o_1}, \bar k'_{\s_2,\o_2},
\bar k'_{\s_3,\o_3})
\hat\psi^+_{\kk'_1,\s_1,\o_1}\hat\psi^+_{\kk'_2,\s_2,\o_2}
\hat\psi^-_{\kk'_3,\s_3,\o_3}\hat\psi^-_{\kk'_4,\s_4,\o_4}\d(\sum_i\e_i
(\kk_i+\vec p_{F,\s_i})\Eq(amba7)$$
and
$$
\d_{h}(\bar k'_{\o,\s})={Z_h(\bar k'_{\s,\o})
\over Z_{h-1}(\bar k'_{\s,\o})}(z_h(\bar
k'_{\s,\o})-a_h(\bar k'_{\s,\o}))\quad\Eq(31)$$
$$\l_{h}
(\bar k'_{\s_1,\o_1}, \bar k'_{\s_2,\o_2},
\bar k'_{\s_3,\o_3})=
[\prod_{i=1}^4\sqrt{Z_h(\bar k'_{\s_i,\o_i})
\over Z_{h-1}(\bar k'_{\s_i,\o_i})}]
l_{h}(\bar k'_{\s_1,\o_1}, \bar k'_{\s_2,\o_2},
\bar k'_{\s_3,\o_3})$$
We will call $\d_{h}$ and $\l_h$ {\it running coupling functions}; 
the above procedure allow to write a recursive equation
for them, see \S 5.

Then we write
$$\int
P_{I,Z_{h-1}}(d\psi^{(\le h-1)})\int
P_{II,Z_{h-1}}(d\psi^{(\le h-1)})
\int
P_{I,Z_{h-1}}(d\psi^{(h)})\int
P_{II,Z_{h-1}}(d\psi^{(h)})$$
$$e^{-\LL\hat \VV^{(h)}(\sqrt{Z_{h-1}}\psi^{(\le h)})
-\RR\VV^{(h)}(\sqrt{Z_{h-1}}\psi^{(\le h)})}\Eq(61)$$
and the propagator of 
$P_{\s,Z_{h-1}}(d\psi)$ is 
$$\hat g^h_{\o,\s}(\kk')= H(a_0^2\sin^2 \hat k_\s)
{1\over Z_{h-1}(\bar k'_{\o,\s})}
{\tilde f_h(k_0,\underline k'_\s,\hat k_\s)
\over -i k_0+2\o\cos \hat k_\s\sin\underline k'_\s}\Eq(62)$$
and
$$\tilde f_h(k_0,\underline k'_\s,\hat k_\s)=Z_{h-1}(\bar k'_{\o,\s})
[{C^{-1}_h(k_0,\underline k'_\s,\hat k_\s)\over
Z_{h-1}(\kk')}-{C_{h-1}^{-1}(k_0,\underline k'_\s,\hat k_\s)\over
Z_{h-1}(\bar k'_{\s,\o})}]\Eq(63)$$
with $H(a_0^2\sin^2 \hat k_\s) \tilde f_h(k_0,\underline k'_\s,\hat k_\s)$ having the same
support that $H(a_0^2\sin^2 \hat k_\s) f_h(k_0,\underline k'_\s,\hat k_\s)$. 

We integrate then the field $\psi^h$ and we get

$$e^{-L^2\b E_{h-1}}\int
P_{I,Z_{h-1}}(d\psi^{(\le h-1)})\int
P_{II,Z_{h-1}}(d\psi^{(\le h-1)})
e^{- \VV^{(h-1)}(\sqrt{Z_{h-1}}\psi^{(\le h-1)})}\Eq(64)$$
and the procedure can be iterated.

We will see in the following section that
if the running coupling functions are small
$$\sup_{k\ge h}
\sup_{\bar k_{\s,\o}}
|\d_{k}(\bar k'_{\s,\o})|\le 2|\l|
\qquad\sup_{k\ge h}\sup_{\bar k'_{\s,\o}}
{Z_{k-1}(\bar k'_{\s,\o})\over Z_{k}(\bar k'_{\s,\o})}\le e^{ 2|\l|}$$ 
$$\sup_{k\ge
h}\sup_{\bar k'_{\s_1,\o_1},\bar
k'_{\s_2,\o_2},\bar k'_{\s_3,\o_3}}|
\l_{k}(\bar k'_{\s_1,\o_1},\bar
k'_{\s_2,\o_2},\bar k'_{\s_3,\o_3})|\le 2|\l|\Eq(f)$$
then the effective potential is given by a convergent series. In \S 4 we will show that
up to exponentially small temperatures this is indeed true.

\vskip.5cm
\section(32, Analyticity of the effective potential)
\vskip.5cm
\*
\sub(3.1) {\it Coordinate representation.}
It is convenient to perform bounds to introduce the variables 
$\psi^\e_{\xx,\o,\s}$. We define the fields
$\psi^\e_{\xx,\o,\s}=e^{i\e\o \vec p_{F,\s}\vec x}
\tilde\psi^\e_{\xx,\o,\s}$, or more explicitly

$$\psi^\e_{\xx,\o,I}=e^{i\e\o p_F x_1}
\tilde\psi^\e_{\xx,\o,I}\qquad
\psi^\e_{\xx,\o,II}=e^{i\e\o p_F x_2}
\tilde\psi^\e_{\xx,\o,II}\Eq(vcv)$$
and the propagators of such fields is
$$\tilde g^h_{\o,\s}(\xx-\yy)=
\int d\kk'{1\over Z_{h-1}(\bar k'_{\o,\s})} 
e^{-i\kk'(\xx-\yy)}{H(a_0^2\sin^2 \hat k_\s)
\tilde f_h(k_0,\underline k'_\s,\hat k_\s)
\over -i k_0+\o 2 \sin \underline k'_\s \cos \hat k_\s}\Eq(ba)$$
It is easy to prove, by integration by parts, that
for any integer $N$, for $L\to\io$
$$|\partial^{n_0}_{x_0}\partial^{n_+}_{x_+}\partial^{n_-}_{x_-}
\tilde g^h _{I,\o}(\xx-\yy)|\le {C_{n_0,n_+,n_-,N}\g^{h(1+n_0+n_+)
}\over 1+
[\g^h|d(x_0-y_0)|+
\g^h|x_+-y_+|+|x_--y_-|]^N}\Eq(f1)$$
$$|\partial^{n_0}_{x_0}\partial^{n_+}_{x_+}\partial^{n_-}_{x_-}
\tilde g^h _{II,\o}(\xx-\yy)|\le {C_{n_0,n_+,n_-,N}\g^{h(1+n_0+n_2)}
\over 1+
[\g^h|d(x_0-y_0)|+|x_+-y_+|+\g^h|x_--y_-|]^N}\Eq(f2)$$
where $d(x_0)={\b\over\pi}\sin {x_0\pi\over\b}$.

{\it Proof.} The above formula can be derived by integration by parts;
note that, if for instance $\s=I$
$$\partial_{k_-} {1\over -i k_0+2\o\sin k'_+\cos k_-}=
{1\over (-i k_0+2\o\sin k'_+\cos k_-)^2}2\o \sin k'_+\sin k_-\Eq(abn)$$
which is $O(\g^{-h})$; in the same way the $n$-th derivative 
with respect to $k_-$ is still $O(\g^{-h})$.
On the other hand $\partial^{n_0}_{k_0}\partial^{n_+}_{k_+}$
is bounded by $\g^{-h-n_0 h-n_+ h}$; finally the integration 
gives a volume factor
$\g^{2h}$.
\vskip.5cm
We define
$$W_{2n}^h(\xx_1,...,\xx_{2n})=
{1\over (L^2\b)^{2n}}\sum_{\kk'_1,...,\kk'_{2n}}
e^{-i\sum_{r=1}^{2n}\e_r\kk'_r\xx_r}
\hat W_{2n}^h(\kk'_1,...\kk'_{2n-1})\d(\sum_i\e_i (\kk'_i+\o_i \vec p_{F,\s_i}))
\Eq(82)$$

Hence \equ(3.3aa) can be written as
$$\VV^{(h)}(\psi^{\le h})=\sum_{n=1}^\io\sum_{\o_1,...,\o_{2n}}
\sum_{\s_1,...,\s_{2n}}\sum_{\e_1,...,\e_{2n}}
\int d\xx_1...d\xx_{2n} 
\left[\prod_{i=1}^{2n} 
\tilde\psi^{(\le
h)\e_i}_{\s_i,\xx_i,\o_i}\right]W^{(h)}_{2n}(\xx_1,...,\xx_{2n})\Eq(98)$$

We now discuss the action of the operator $\LL$
and $\RR=1-\LL$ on the effective potential
in the $x$-space representation. Noting that from \equ(82),
if $\e_1=\e_2=-\e_3=-\e_4=+$
$$W_4^h(\xx_1,\xx_2,\xx_3,\xx_4)=
e^{i\vec x_4(\o_1 \vec p_{F,\s_1}+\o_2 \vec p_{F,\s_2}-\o_3
\vec p_{F,\s_3}
-\o_4 \vec p_{F,\s_4})}
\tilde W_4^h(\xx_1-\xx_4,\xx_2-\xx_4,\xx_3-\xx_4)\Eq(85)$$
we can write the action of $\RR$ \equ(loc2) as
$$\RR \int \prod_{i=1}^4 d\xx_i
\prod_{i=1}^4\tilde\psi^{\e_i}_{\xx_i,\s_i,\o_i}
e^{i\vec x_4
(\o_1 \vec p_{F,\s_1}+\o_2 \vec 
p_{F,\s_2}-\o_3 \vec p_{F,\s_3}
-\o_4 \vec p_{F,\s_4})}\tilde 
W_4^h$$
$$=\int \prod_{i=1}^4 d\xx_i
\prod_{i=1}^4\tilde\psi^{\e_i}_{\xx_i,\s_i,\o_i}
e^{i\xx_4(\o_1 p_{F,\s_1}+\o_2 p_{F,\s_2}-\o_3 p_{F,\s_3}
-\o_4 p_{F,\s_4})}$$
$$[\tilde W_4^h(\xx_1-\xx_4,\xx_2-\xx_4,\xx_2-\xx_4)
-\d(x_{0,1}-x_{0,4})
\d(x_{0,2}-x_{0,4})\d(x_{0,3}-x_{0,4})
\d(\underline x_{\s_1,1}-\underline x_{\s_1,4})$$
$$\d(\underline x_{\s_2,2}-\underline x_{\s_2,4})
\d(\underline x_{\s_3,3}-\underline x_{\s_3,4})
\int dt_{0,1}dt_{0,2}dt_{0,3} 
d \underline t_{1,\s_1}d \underline t_{2,\s_2}d \underline t_{3,\s_3}
\tilde W^{(h)}_4({\bf t}_{1},{\bf t}_{2}, 
{\bf t}_{3})\Eq(loc7)$$
where $\int d\xx={\b\over M}\sum_{\xx\in\L}$ and
${\bf t}_i=(t_{0,i},\underline
t_{\s,i}, \hat t_{\s,i})$ where
$\underline t_{I,i}=t_{+,i}; 
\underline t_{II,i}=t_{-,i}$ and $\hat t_{I,i}=t_{-,i}; 
\hat t_{II,i}=t_{+,i}$.

On the other hand we can equivalently write 
the $\RR$ operation as acting on the fields, and such 
two representations
of the $\RR$ operation will be used in the following.
It holds that, by simply integrating the deltas in \equ(loc7)
$$\RR \int [\prod_{1=1}^4 d\xx_i]
W^h_4(\{\xx\}) \tilde\psi^+_{\xx_1,\s_1,\o_1}
\tilde\psi^+_{\xx_{2},\s_2,\o_2}
\tilde\psi^-_{\xx_{3},\s_3,\o_3}
\tilde\psi^-_{\xx_{4},\s_4,\o_4}=
\int [\prod_{1=1}^4 d\xx_i] W^h_4(\{\xx\})\times$$
$$[D^+_{\xx_1,\bar\xx_{4,\s_1},\s_1,\o_1}
\tilde\psi^+_{\xx_{2},\s_2,\o_2}\tilde
\psi^-_{\xx_{3},\s_3,\o_3}
\tilde\psi^-_{\xx_{4},\s_4,\o_4}
+\tilde\psi^+_{\bar \xx_4,\s_1,\o_1}
D^+_{\xx_2,\bar\xx_{4,\s_2},\s_2,\o_2}
\tilde\psi^-_{\xx_{3},\s_3,\o_3}
\tilde\psi^-_{\xx_{4},\s_4,\o_4}+$$
$$\tilde\psi^+_{\bar \xx_4,\s_1,\o_1}
\tilde\psi^+_{\bar\xx_4,\s_2,\o_2}
D^-_{\xx_3,\bar\xx_{4,\s_3},\s_3,\o_3}
\tilde\psi^-_{\xx_{4},\s_4,\o_4}\Eq(f7)$$
where $\bar\xx_{4,\s_i}
=(x_{0,4},x_{+,4},x_{-,i})$ if $\s_i=I$ 
and
$\bar\xx_{4,\s_i}=
(x_{0,4},x_{+,i},x_{-,4})$ if $\s_i=II$; 
moreover
$$D^\e_{\xx_i,\bar\xx_{4,\s_i},\s_i,\o_i}
=\tilde\psi^\e_{\xx_i,\s_i,\o_i}
-\tilde\psi^\e_{\bar\xx_{4,\s_i},\s_i,\o_i}\Eq(bnm)$$
This means that the action of the renormalization operator $\RR$ 
can be seen as the replacement of a 
$\psi^{\e(\le h)}$ field  with a
$D^{\e(\le h)}_{\xx_i,\bar\xx_{4,\s_i},\s_i,\o_i}$ field 
and some of the other $\tilde\psi^{(\le h)}$ fields are
``translated'' in the localization point. 
The field $D^{\e(\le h)}_{\xx_i,\bar\xx_{4,\s_i},\s_i,\o_i}$ 
is antiperiodic
in the time components of $\xx_{i}$, and $\bar\xx_{4}$.
We can write $D^{\e(\le h)}$ as sum of two terms 
(if $\s_i=I$ for instance):
$$D^{\e(\le h)}_{\xx_i,\bar\xx_{4,\s_i},\s_i,\o_i}=
[\tilde\psi^\e_{\xx_i,I}-
\tilde\psi^\e_{x_{0,4},x_{+,i},x_{-,i},I}]+
[\tilde\psi^\e_{x_{0,4},x_{+,i},x_{-,i},I}
-\tilde\psi^\e_{x_{0,4},x_{+,4},x_{-,i},I}]\Eq(f1)$$
and the second addend can be written as, for $L\to\io$
$$\tilde\psi^\e_{x_{0,4},x_{+,i},x_{-,i},I}
-\tilde\psi^\e_{x_{0,4},x_{+,4},x_{-,i},I}=
(x_{+,i}-x_{+,4})\int_0^1
 dt \partial_{x_{+}}
\tilde\psi^\e_{I,\o,x_{0,4},x_{+,i}-t(x_{+,i}-x_{+,4}),x_{-,i}} \Eq(f10)$$
and $x_{+,i}-t(x_{+,i}-x_{+,4})\equiv x_{+,i,4}(t)$
is called {\it interpolated point}.

This means
that it is
dimensionally equivalent to the product of the "zero" 
$(x_{+,i}-x_{+,4})$ and
the derivative of the field, so that the bound of its contraction with another
field variable on a scale $h'<h$ will produce a ``gain'' $\g^{-(h-h')}$,
see \equ(f1),\equ(f2),  with
respect to the contraction of $\tilde\psi^{(\le h)\sigma}_{\xx,\o}$.
Similar considerations can be repeated for the first addend of \equ(f1);
some care has to be done as $\b$ is finite, 
and we refer \S 3.5 of [BM].
If there are two external lines.

$$\RR \int d\xx_1 d\xx_2 W^h_2(\xx_1,\xx_2)
\tilde\psi^+_{\xx_1,\s_1,\o_1}
\tilde\psi^-_{\xx_{2},\s_2,\o_2}=\int d\xx_1 d\xx_2 W^h_2(\xx_1,\xx_2)
\tilde\psi^+_{\xx_1,\s_1,\o_1}
T^-_{\xx_2,\bar \xx_{1,\s_2},\s_2,\o_2}\Eq(f8)$$
where
$$T^\e_{\xx_2,\bar\xx_{1,\s_2},\s_2,\o}
=\tilde\psi^\e_{\xx_2,\s_2,\o}-
\tilde\psi^{\e}_{\bar\xx_{1,\s_2},\o,\s_2}-(x_{0,2}-x_{0,1})
\partial_{0}\tilde\psi^{\e}_{\bar\xx_{1,\s_2},\o,\s_2}
-(\underline x_{\s,2}-\underline x_{1,\s})
\partial_{\s}\tilde\psi^{-}_{\bar\xx_{1,\s_2},\o,\s_2}\Eq(f11)$$
and $\partial_\s=\partial_{x_+}$ if $\s=I$ and $\partial_\s=\partial_{x_-}$
if $\s=II$.
In this case the "gain"
produced by the $\RR$ operation is $\g^{-2(h-h')}$.

We can write the local part of the effective potential
\equ(loc3) in the following way
$$\LL V^h=\sum_{\s_1=\s_2}\int d\xx_1 d\tilde x_{\s,2}
[\o\d_{h,\o}((\hat x_{1}-\hat x_2)_{\s_1})
e^{i\vec x_2(\o_1 \vec p_{F,\s_1}-\o_2 \vec 
p_{F,\s_2})}
\tilde\psi^{+}_{\o,\s_1;\xx_1}
\partial_\s\tilde\psi^{-}_{\o,\s_2;\bar\xx_{1,\s_2}}+\Eq(f5)$$
$$\sum_{\s_1,..,\s_4=I,II}\int d\xx_4 d\tilde x_{\s_1,1} 
d\tilde x_{\s_2,2} d\tilde x_{\s_3,3}
\l_{h;\o_1,..\o_4}((\hat x_{1}-\hat x_4)_{\s_1},
(\hat x_{2}-\hat x_4)_{\s_1},(\hat x_{3}
-\hat x_4)_{\s_3})$$
$$e^{i\vec x_4(\o_1 \vec p_{F,\s_1}+\o_2 \vec 
p_{F,\s_2}-\o_3 \vec p_{F,\s_3}
-\o_4 \vec p_{F,\s_4})}\tilde\psi^+_{\bar\xx_{4,\s_1},\s_1,\o_1}
\tilde\psi^+_{\bar\xx_{4,\s_2},\s_2,\o_2}
\tilde\psi^-_{\bar\xx_{4,\s_3},\s_3,\o_3}
\tilde\psi^-_{\xx_4,\s_4,\o_4}$$
where 
$\d_h(\xx)$ is the Fourier transform of $\d_h(\hat k_\s)2\cos\hat
k_\s$ with respect to $\hat k_\s$ and
$(\hat x_{i}-\hat x_j)_{\s}
=x_{-,i}-x_{-,j}$ if $\s=I$ and
$(\hat x_{i}-\hat x_j)_{\s}
=x_{+,i}-x_{+,j}$
if $\s=II$; moreover $\tilde x_{\s_i}=x_-$ if $\s=I$ and  
$\tilde x_{\s_i}=x_+$ if $\s=II$.
\*
\sub(3.2) {\it Tree expansion.}
By using iteratively the ``single scale expansion'' 
we can write the effective
potential $\VV^{(h)}(\psi^{(\le h)})$, for $h\le 0$,
in terms of a {\it tree expansion}. For a tutorial introduction
to the tree formalism we will refer to the review [GM].
  
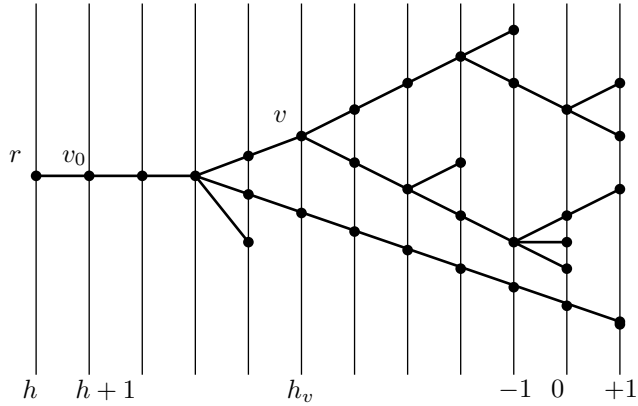
\insertplot{300pt}{150pt}%
{\ins{30pt}{85pt}{$r$}\ins{50pt}{85pt}{$v_0$}\ins{130pt}{100pt}{$v$}%
\ins{35pt}{-2pt}{$h$}\ins{55pt}{-2pt}{$h+1$}\ins{135pt}{-2pt}{$h_v$}%
\ins{215pt}{-2pt}{$-1$}\ins{235pt}{-2pt}{$0$}\ins{255pt}{-2pt}{$+1$}}%
{fig51}{\eqg(1)}

\vglue.5truecm

We need some definitions and notations.
 
\0 1) Let us consider the family of trees which can be constructed
by joining a point $r$, the {\it root}, with an ordered set of $n\ge 1$
points, the {\it endpoints} of the {\it unlabeled tree} (see Fig. 1),
so that $r$ is not a branching point. $n$ will be called the
{\it order} of the unlabeled tree and the branching points will be
called
the {\it non trivial vertices}.
The unlabeled trees are partially ordered from the root to the
endpoints in
the natural way; we shall use the symbol $<$ to denote the partial
order.
 
Two unlabeled trees are identified if they can be superposed by a
suitable
continuous deformation, so that the endpoints with the same index
coincide.
It is then easy to see that the number of unlabeled trees with $n$
end-points
is bounded by $4^n$.
 
We shall consider also the {\it labeled trees} (to be called simply
trees in
the following); they are defined by associating some labels with the
unlabeled
trees, as explained in the following items.
 
\0 2) We associate a label $h\le -1$ with the root and we denote
$\TT_{h,n}$ the
corresponding set of labeled trees with $n$ endpoints. Moreover, we
introduce
a family of vertical lines, labeled by an integer taking values in
$[h,1]$, and we represent any tree $\t\in\TT_{h,n}$ so that, if $v$ is
an
endpoint or a non trivial vertex, it is contained in a vertical line
with
index $h_v>h$, to be called the {\it scale} of $v$, while the root is
on the
line with index $h$. There is the constraint that, if $v$ is an
endpoint,
$h_v>h+1$.
 
The tree will intersect in general the vertical lines in set of
points different from the root, the endpoints and the non trivial
vertices;
these points will be called {\it trivial vertices}. The set of the {\it
vertices} of $\t$ will be the union of the endpoints, the trivial
vertices
and the non trivial vertices.
Note that, if $v_1$ and $v_2$ are two vertices and $v_1<v_2$, then
$h_{v_1}<h_{v_2}$. We will call $s_v$ the number of subtrees coming out
from $v$.
 
Moreover, there is only one vertex immediately following
the root, which will be denoted $v_0$ and can not be an endpoint;
its scale is $h+1$.

\0 3) To each end-point of scale $+1$ we associate $\VV$ (1.11).
With each endpoint $v$
of
scale $h_v\le 0$ we associate one of the two terms appearing
in \equ(f5), with coupling $\l_{h_v-1}$ or $\d_{h_v-1}$.
Moreover, we impose the constraint that, if $v$ is an endpoint 
and $h_v\le 0$,
$h_v=h_{v'}+1$, if $v'$ is the non trivial vertex immediately preceding
$v$.

\0 4) We introduce a {\it field label} $f$ to distinguish the field
variables
appearing in the terms associated with the endpoints as in item 3);
the set of field labels associated with the endpoint $v$ will be called
$I_v$.
Analogously, if $v$ is not an endpoint, we shall
call $I_v$ the set of field labels associated with the endpoints
following
the vertex $v$; $\xx(f)$, $\e(f)$ and $\o(f)$ will denote the
space-time
point, the $\e$ index and the $\o$ index, respectively, of the
field variable with label $f$.
 
If $h_v\le 0$, one of the field variables belonging to $I_v$ carries also a
derivative $\dpr_{\s}$ if the corresponding local
term is of type $\d$, see \equ(f5).
Hence we can associate with each field label $f$ an integer $m(f)\in\{0,1\}$,
denoting the order of the derivative. 
 
\*
If $h\le -1$,
the effective potential can be written in the following way:
$$\VV^{(h)}(\psi^{(\le h)}) + L\b E_{h+1}=
\sum_{n=1}^\io\sum_{\t\in\TT_{h,n}}
\VV^{(h)}(\t,\psi^{(\le h)})\;,\Eq(3.29)$$
where, if $v_0$ is the first vertex of $\t$ and $\t_1,..,\t_s$
($s=s_{v_0}$)
are the subtrees of $\t$ with root $v_0$,\\
$\VV^{(h)}(\t,\psi^{(\le h)})$ is defined inductively by the relation
$$\eqalign{
&\qquad \VV^{(h)}(\t,\psi^{(\le h)})=\cr
&{(-1)^{s+1}\over s!} \EE^T_{h+1}[
\bar\VV^{(h+1)}(\t_1,\psi^{(\le h+1)});..;
\bar\VV^{(h+1)}(\t_{s},\psi^{(\le
h+1)})]\;,\cr}\Eq(3.30)$$
and $\bar\VV^{(h+1)}(\t_i,\psi^{(\le h+1)})$
 
\0 a) is equal to $\RR\VV^{(h+1)}(\t_i,\psi^{(\le h+1)})$ if
the subtree $\t_i$ is not trivial, 
with $\RR$ defined as acting on kernels
according to \equ(loc7) and its analogous for $n=1$;
 
\0 b) if $\t_i$ is trivial and $h< -1$, it
is equal to $\LL\VV^{(h+1)}(\psi^{(\le h+1)})$ \equ(f5) or,
if $h=-1$, to $\VV$.
 
\acapo
$\EE^T_{h+1}$ denotes the truncated expectation with respect
to the measure $\prod_{\s=I,II}P_{\s,Z_{h}}(d\psi^{(h+1)})$,
that is
$$\EE^T_{h+1}(X_1;\ldots;X_p)\={\dpr^p\over\dpr\l_1\ldots\dpr\l_p}
\left.\log\int\prod_{\s=I,II}P_{\s,Z_{h}}(d\psi^{(h+1)})
e^{\l_1 X_1+ \cdots\l_pX_p}\right|_{\l_i=0}.\Eq(3.31)$$

We write \equ(3.30) in a more explicit way.
If $h=-1$, the r.h.s. of
\equ(3.30) can be written in the following way. Given
$\t\in\TT_{-1,n}$, there are $n$ endpoints of scale $1$ 
and only another one
vertex, $v_0$, of scale $0$; let us call $v_1,\ldots, v_n$ the endpoints. We
choose, in any set $I_{v_i}$, a subset $Q_{v_i}$ and we define $P_{v_0}=\cup_i
Q_{v_i}$. We have
$$\VV^{(-1)}(\t, \psi^{(\le -1)})=\sum_{P_{v_0}}
\VV^{(-1)}(\t,P_{v_0})\;,\Eq(3.31a)$$
$$\VV^{(-1)}(\t,P_{v_0})= \int d\xx_{v_0}
\tilde\psi^{\le -1} (P_{v_0}) K_{\t,P_{v_0}}^{(0)}(\xx_{v_0})\;,
\Eq(3.31b)$$
$$K_{\t,P_{v_0}}^{(0)}(\xx_{v_0})={1\over n!} \EE^T_{0}[
\tilde\psi^{(0)}(P_{v_1}\bs Q_{v_1}),\ldots,
\tilde\psi^{(0)}(P_{v_n}\bs Q_{v_n})]
\prod_{i=1}^n K^{(1)}_{v_i}(\xx_{v_i})\;,\Eq(3.31c)$$
where we use the definitions
$$\tilde\psi^{(\le h)}(P_v)= \prod_{f\in P_v}
\hat\dpr_{\s(f)}^{m(f)}
\tilde\psi^{(\le h)\e(f)}_{\xx(f)}\virg h\le -1\;,\Eq(3.34)$$
$$\tilde\psi^{(0)}(P_v) =\prod_{f\in P_v} \tilde\psi^{(0)\s(f)}_{\xx(f)}
\;,\Eq(3.31d)$$
$$K^{(1)}_{v_i}(\xx_{v_i})= e^{i\sum_{f\in
I_{v_i}}\e_f\vec x(f)\o(f)\vec p_{F,\s(f)}
}\l\qquad \xx_{v_i}=\xx\Eq(3.21)$$
It is not hard to see that, by iterating the previous procedure, one gets for
$V^{(h)}(\t,\psi^{(\le h)})$, for any $\t\in\TT_{h,n}$, the representation
described below.
We associate with any vertex $v$ of the tree a subset $P_v$ of $I_v$, the {\it
external fields} of $v$. These subsets must satisfy various constraints. First
of all, if $v$ is not an endpoint and $v_1,\ldots,v_{s_v}$ are the vertices
immediately following it, then $P_v \subset \cup_i P_{v_i}$; if $v$ is an
endpoint, $P_v=I_v$. We shall denote $Q_{v_i}$ the intersection of $P_v$ and
$P_{v_i}$; this definition implies that $P_v=\cup_i Q_{v_i}$. The subsets
$P_{v_i}\bs Q_{v_i}$, whose union ${\cal I}_v$ will be made, by definition, of
the {\it internal fields} of $v$, have to be non empty, if $s_v>1$.
Moreover, we associate with any $f\in {\cal I}_v$ a scale label $h(f)=h_v$.
Given $\t\in\TT_{h,n}$, there are many possible choices of the subsets $P_v$,
$v\in\t$, compatible with all the constraints; we shall denote $\PP_\t$ the
family of all these choices and $\bP$ the elements of $\PP_\t$. 
 
Then we can write
$$\VV^{(h)}(\t,\psi^{(\le h)})=\sum_{\bP\in\PP_\t}
\VV^{(h)}(\t,\bP)\;.\Eq(3.32)$$
$\VV^{(h)}(\t,\bP)$ can be represented as
$$\VV^{(h)}(\t,\bP)=\int d\xx_{v_0} \tilde\psi^{(\le h)}
(P_{v_0}) K_{\t,\bP}^{(h+1)}(\xx_{v_0})\;,\Eq(3.33)$$
with $K_{\t,\bP}^{(h+1)}(\xx_{v_0})$ defined inductively (recall that
$h_{v_0}=h+1$) by the equation, valid for any $v\in\t$ which is not an
endpoint,
$$K_{\t,\bP}^{(h_v)}(\xx_v)={1\over s_v !}
\prod_{i=1}^{s_v} [K^{(h_v+1)}_{v_i}(\xx_{v_i})]\;
\;\EE^T_{h_v}[ \tilde\psi^{(h_v)}(P_{v_1}\bs Q_{v_1}),\ldots,
\tilde\psi^{(h_v)}(P_{v_{s_v}}\bs Q_{v_{s_v}})]\;,\Eq(3.35)$$
where $\tilde\psi^{(h_v)}(P_v)$ is defined as in \equ(3.34), with $(h_v)$
in place of $(\le h_v)$, if $h_v\le -1$, while, if 
$h_v=0$, it is defined as in \equ(3.31d).

Moreover, if $v$ is an endpoint and $h_v=0$, $K^{(1)}_{v}(\xx_{v})$ 
is given by \equ(3.21), otherwise, see \equ(f5)
$$K^{(h_v)}_{v}(\xx_{v})= \cases{ 
l_{h_v-1}(\vec x_1,\vec x_2,\vec x_3,\vec x_4)& if $v$ 
is of type $\l$,\cr
d_{h_v-1}(\vec x_1,\vec x_2) & if $v$ is of type $z$,}\Eq(3.37)$$
where 
$$l_{h_v-1}(\vec x_1,\vec x_2,\vec x_3,\vec x_4)=$$
$$e^{i\vec x_4(\e_1\o_1 \vec p_{F,\s_1}+\e_2\o_2 \vec p_{F,\s_2}
-\o_3 \vec p_{F,\s_3}
-\e_4\o_4 \vec p_{F,\s_4})}
\l_{h_v-1,\o_1,..,\o_4}((\hat x_1-\hat x_4)_{\s_1},
(\hat x_2-\hat x_4)_{\s_2},(\hat x_3-\hat x_4)_{\s_3})$$
$$d_{h_v-1,\o}(\vec x_1,\vec x_2)=
e^{i\vec x_2(\o_1 \vec p_{F,\s_1}-\e_2\o_2 \vec p_{F,\s_2})}
\d_{h_v-1}((\hat x_1-\hat x_2)_{\s_1})$$
If $v$ is not an endpoint,
$$K^{(h_v+1)}_{v_i}(\xx_{v_i}) = \RR K_{\t_i,\bP^{(i)},\O^{(i)}}^{(h_v+1)}
(\xx_{v_i})\;,\Eq(3.37a)$$
where $\t_i$ is the subtree of $\t$ starting from $v$ and passing through
$v_i$ (hence with root the vertex immediately preceding $v$), $\bP^{(i)}$ and
is the restrictions to $\t_i$ of $\bP$. The action of
$\RR$ is defined using the representation \equ(loc7) of the $\RR$ operation.

\*

\equ(3.32) is not the final form of our expansion, since we further decompose
$\VV^{(h)}(\t,\bP)$, by using the following representation of the truncated
expectation in the r.h.s. of \equ(3.35). Let us put $s=s_v$, $P_i\=P_{v_i}\bs
Q_{v_i}$; moreover we order in an arbitrary way the sets $P_i^\pm\=\{f\in
P_i,\s(f)=\pm\}$, we call $f_{ij}^\pm$ their elements and we define
$\xx^{(i)}=\cup_{f\in P_i^-}\xx(f)$, $\yy^{(i)}=\cup_{f\in P_i^+}\xx(f)$,
$\xx_{ij}=\xx(f^-_{i,j})$, $\yy_{ij}=\xx(f^+_{i,j})$. Note that $\sum_{i=1}^s
|P_i^-|=\sum_{i=1}^s |P_i^+|\=n$, otherwise the truncated expectation
vanishes. A couple $l\=(f^-_{ij},f^+_{i'j'})\=(f^-_l,f^+_l)$ will be called a
line joining the fields with labels $f^-_{ij},f^+_{i'j'}$ and sector indices
$\o^-_l=\o(f^-_l)$, $\o^+_l=\o(f^+_l)$ and connecting the points
$\xx_l\=\xx_{i,j}$ and $\yy_l\=\yy_{i'j'}$, the {\it endpoints} of $l$.
Moreover, we shall put $m_l=m(f^-_l)+m(f^+_l)$ and, if $\o^-_l=\o^+_l$,
$\o_l\=\o^-_l=\o^+_l$. A similar definition is repeated for $\s$.
Then, it is well known (see [Le], [BM], [GM] for
example) that, up to a sign, if $s>1$,
$$\EE^T_{h}(\tilde\psi^{(h)}(P_1),...,\tilde\psi^{(h)}(P_s))=
\sum_{T}\prod_{l\in T}  
\tilde g^{(h)}_{\o^-_l,\s^-_l}(\xx_l-\yy_l)
\d_{\o^-_l,\o^+_l}\d_{\s^-_l,\s^+_l}
\int dP_{T}(\tt) \det G^{h,T}(\tt)\Eq(3.38)$$
where
$T$ is a set of lines forming an {\it anchored tree graph} between the
clusters of points $\xx^{(i)}\cup\yy^{(i)}$, that is $T$ is a set of lines,
which becomes a tree graph if one identifies all the points in the same
cluster. Moreover $\tt=\{t_{i,i'}\in [0,1], 1\le i,i' \le s\}$, $dP_{T}(\tt)$
is a probability measure with support on a set of $\tt$ such that
$t_{i,i'}=\uu_i\cdot\uu_{i'}$ for some family of vectors $\uu_i\in \RRR^s$ of
unit norm. Finally $G^{h,T}(\tt)$ is a $(n-s+1)\times (n-s+1)$ matrix, whose
elements are given by $G^{h,T}_{ij,i'j'}=t_{i,i'} \hat\dpr_{\s(f^-_{ij})}
^{m(f^-_{ij})}\hat\dpr_{\s(f^+_{ij})}
^{m(f^+_{ij})}
\tilde g^{(h)}_{\o_l}(\xx_{ij}-\yy_{i'j'})\d_{\o^-_l,\o^+_l}
\d_{\s^-_l,\s^+_l}$ with
$(f^-_{ij}, f^+_{i'j'})$ not belonging to $T$.

In the following we shall use \equ(3.38) even for $s=1$, when $T$ is empty, by
interpreting the r.h.s. as equal to $1$, if $|P_1|=0$, otherwise as equal to
$\det G^{h}=\EE^T_{h}(\tilde\psi^{(h)}(P_1))$.
\*

If we apply the expansion \equ(3.38) in each non trivial vertex of $\t$, we
get an expression of the form
$$\VV^{(h)}(\t,\bP) = \sum_{T\in {\bf T}} \int d\xx_{v_0}
\tilde\psi^{(\le h)}(P_{v_0})
W_{\t,\bP,T}^{(h)}(\xx_{v_0})\=
\sum_{T\in {\bf T}} \VV^{(h)}(\t,\bP,T)\;,\Eq(3.38a)$$
where ${\bf T}$ is a special family of graphs on the set of points
$\xx_{v_0}$, obtained by putting together an anchored tree graph $T_v$ for
each non trivial vertex $v$. Note that any graph $T\in {\bf T}$ becomes a tree
graph on $\xx_{v_0}$, if one identifies all the points in the sets $x_v$, for
any vertex $v$ which is also an endpoint.

We are writing the $\RR$ operation as acting on the kernels, according
to \equ(loc7) and its analogous for $n=1$. 
Such representation for the $\RR$ operation
is however not suitable 
to "gain" the convergence factor $\g^{-(h-h')}$,
or $\g^{-2(h-h')}$,
for which is much more convenient
representation of $\RR$
in \equ(f7), \equ(f8). However if we write simply all the
$\RR$ operations as in \equ(f7), \equ(f8)
one gets possibly factors $(\xx_i-\xx_j)^{\a_n}$
with $\a_n=O(n)$, which when integrated
give $O(n!)$ terms. One has to proceed
in a more subtle way starting 
from the vertices of $\t$ closest
to the root from which the $\RR$ operation is non trivial,
and writing $\RR$ as in
\equ(f7),\equ(f8) leaving all the other $\RR$
operation as in \equ(loc7). One distributes the "zero" 
along a path connecting a family of end points, and
from \equ(loc7)
$(\xx_i-\xx_j)\RR\tilde W^h_{4}=
(\xx_i-\xx_j)\tilde W^h_{4}$, if $\xx_i,\xx_j$ are two coordinates of
$\tilde W^h_4$ and $\RR\tilde W^h_{4}$ is the term in square
brakets in the l.h.s. of \equ(loc7); an analogous property
holds for $\RR\tilde W^h_{2}$.
There
are same technical complications in implementing this idea,
which are discussed in [BM] (see also [BoM]), 
\S 3.2, \S 3.3 for a different model, but the adapting
of such argument to the present case is straightforward.
We obtain, in the $L\to\io$ limit

$$\eqalign{
V^{(h)}(\t,\bP)&=
\sum_{T\in {\bf T}} \sum_{\a\in A_T}
\int d\xx_{v_0} \sqrt{Z_{h}}^{|P_{v_0}|}
W_{\t,\bP,T,\a}(\xx_{v_0})\;\cdot\cr
&\cdot\; \Big\{ \prod_{f\in P_{v_0}}
[\hat\partial_{j_\a(f)}^{q_\a(f)}\psi]^{(\le h)\s(f)}_{\xx_\a(f),\o(f)}\Big\}
\;,\cr}\Eq(3.80)$$
where
$$\eqalign{
&W_{\t,\bP,T,\a}(\xx_{v_0})=
\Big[\prod_{v\,\hbox{\ottorm not e.p.}}
\Big(Z_{h_v}/Z_{h_v-1}\Big)^{|P_v|/2}\Big]\;\cdot\cr
&\cdot \Big[\prod_{i=1}^n d_{j_\a(v^*_i)}^{b_\a(v^*_i)}(\xx_i,\yy_i)
K^{h_i}_{v^*_i}(\xx_{v^*_i})\Big]
\Big\{\prod_{v\,\hbox{\ottorm not e.p.}}{1\over s_v!} \int
dP_{T_v}(\tt_v) \;\cdot\cr
&\cdot\; \det G_\a^{h_v,T_v}(\tt_v)
\Big[\prod_{l\in T_v} \bar\partial^{q_\a(f^-_l)}_{j_\a(f^-_l)}
\bar\partial^{q_\a(f^+_l)}_{j_\a(f^+_l)} [d^{b_\a(l)}_{j_\a(l)}(\xx_l,\yy_l)
\hat\dpr^{m_l} g^{(h_v)}_{\s^-_l, \o^-_l;
\s^+_l,\o^+_l}(\xx_l-\yy_l)]\Big]\Big\}\;,\cr}
\Eq(3.81)$$
where:

\*
\0 1) $\bP$ is the set of $\{P_v\}$;

\0 2) ${\bf T}$ is the set of the tree graphs on $\xx_{v_0}$, obtained by
putting together an anchored tree graph $T_v$ for each non trivial 
vertex $v$;

\0 3) $A_T$ is a
set of indices which allows to distinguish the different terms produced by
the non trivial $\RR$ operations and the iterative decomposition of the zeros;
$v^*_1,\ldots,v^*_n$ are the endpoints of $\t$, $f^-_l$ and $f^+_l$ are the
labels of the two fields forming the line $l$, ``e.p.''  is an
abbreviation of ``endpoint''. 

\0 4)
$G_\a^{h_v,T_v}(\tt_v)$ is obtained from the matrix $G^{h_v,T_v}(\tt_v)$,
associated with the vertex $v$ and $T_v$, by substituting
$G^{h_v,T_v}_{ij,i'j'}=t_{v,i,i'} \hat\dpr_{\underline x_{\s(f^-_{ij})}}
^{m(f^-_{ij})}\hat\dpr_{\underline x_{\s(f^+_{i'j'})}}
^{m(f^+_{i'j'})}
g^{(h_v)}_{\s^-_l ,\o^-_l;\s^+_l,\o^+_l}(\xx_{ij}-\yy_{i'j'})$ with
$$G^{h_v,T_v}_{\a,ij,i'j'}=t_{v,i,i'}
\bar\partial_{j_\a(f^-_{ij})}^{q_\a(f^-_{ij})}
\bar\partial_{j_\a(f^+_{ij})}^{q_\a(f^+_{ij})}
\hat\dpr_{\underline x_{\s(f^-_{ij})}}
^{m(f^-_{ij})}\hat\dpr_{\underline x_{\s(f^+_{i'j'})}}
^{m(f^+_{i'j'})}
g^{(h_v)}_{\s^-_l ,\o^-_l;\s^+_l,\o^+_l}(\xx_{ij}-\yy_{i'j'})\;.\Eq(3.82)$$

\0 5)$\bar\partial_j^q$, $q=0,1,2$, are discrete derivatives
or operators dimensionally equivalent to derivatives, due to the presence of
the lattice and the fact that $\b$ is finite, see [BM] \S 3.
Morever
$\bar\partial^0_j$ denotes the identity and $j=0,+,-$.
According to \equ(f10), \equ(f11)
if $\s(f)=I$ then in 
$\bar\partial_{j(f)}^q$ one has $j(f)=0,+$ and if 
$\s(f)=II$ then $j(f)=0,-$.

\0 6) $d_0(\xx_l-\yy_l)={\b\over\pi}\sin{\pi\over\b}(x_{0,l}-y_{0,l})$ and
$d_i(\xx_l-\yy_l)=(x_{i,l}-y_{i,l})$, $i=\pm $
are the "zeros" produced by the $\RR$ operation, see 
\equ(f10),\equ(f11). Finally by construction $b_a(l)\le 2$.

\0 7) The factors ${Z_{h-1}\over Z_h}$
are functions of the coordinates, and such dependence is
not explicitly written.

Of course the coefficients $b_\a$ and $q_\a$ are not independent,
and, by the definition of $\RR$ (see the discussion
after \equ(f10)) it holds
for any $\a\in A_T$, the following inequality 
$$\Big[\prod_{f\in I_{v_0}} \g^{h_\a(f) q_\a(f)} \Big]
\Big[\prod_{l\in T} \g^{-h_\a(l) b_\a(l)} \Big] \le
\prod_{v\,\hbox{\ottorm not e.p.}} \g^{-z(P_v)}\;,\Eq(3.83)$$
where $h_\a(f)=h_{v_0}-1$ if $f\in P_{v_0}$, otherwise it is the scale of
the vertex where the field with label $f$ is contracted;
$h_\a(l)=h_v$, if
$l\in T_v$ and
$$z(P_v)=\cases{
1 & if $|P_v|=4$,\cr
2 & if $|P_v|=2$;\cr
0 & otherwise.\cr}\Eq(3.84)$$

It holds

$$\eqalign{
&|\det G_\a^{h_v,T_v}(\tt_v)|
\le 
C^{\sum_{i=1}^{s_v}|P_{v_i}|-|P_v|-2(s_v-1)}\;\cdot\cr
&\cdot\;
\g^{{h_v\over 2}\left(\sum_{i=1}^{s_v}|P_{v_i}|-|P_v|-2(s_v-1)\right)}
\g^{h_v  \sum_{i=1}^{s_v}\left[  q_\a(P_{v_i}\bs Q_{v_i})+ m(P_{v_i}\bs Q_{v_i})
\right] }\;\cdot\cr
&\cdot\; \g^{-h_v \sum_{l\in T_v}\left[ q_\a(f^+_l)+ q_\a(f^-_l)+
m(f^+_l)+m(f^-_l)\right]}\;.\cr}\Eq(3.94)$$

This follows
from the well known {\it Gram-Hadamard inequality},
see
also [Le],[BM],[GM], stating
that, if $M$ is a square matrix with elements $M_{ij}$ of the form
$M_{ij}=<A_i,B_j>$, where $A_i$, $B_j$ are vectors in a Hilbert space
with
scalar product $<\cdot,\cdot>$, then
$$|\det M|\le \prod_i ||A_i||\cdot ||B_i||\;.\Eq(3.49)$$
where $||\cdot||$ is the norm induced by the scalar product.
 
In our case it can be shown that
$$\eqalign{&G^{h_v,T_v}_{\a, ij,i'j'}=t_{i,i'}
\bar\partial_{j_\a(f^-_{ij})}^{q_\a(f^-_{ij})}
\bar\partial_{j_\a(f^+_{ij})}^{q_\a(f^+_{ij})}
\hat\dpr
^{m(f^-_{ij})}\hat\dpr
^{m(f^+_{i'j'})}
g^{(h_v)}_{\o_l,\s_l}(\xx_{ij}-\yy_{i'j'})=\cr
&=<\uu_i\otimes A^{(h_v)}_{\xx(f^-_{ij}),\o_l,\s_l},
\uu_{i'}\otimes B^{(h_v)}_{\xx(f^+_{i'j'}),\o_l,\s_l}>\;,\cr}\Eq(3.50)$$
where $\uu_i\in \RRR^s$, $i=1,\ldots,s$, are the vectors such that
$t_{i,i'}=\uu_i\cdot\uu_{i'}$, and $A^{(h_v)}_{\xx(f^-_{ij}),\o_l,\s_l}$,
$B^{(h_v)}_{\xx(f^+_{i'j'}),\o_l,\s_l}$ are such that (in the case $q=m=0$
for simplicity):
$$\eqalign{
g^{(h_v)}_{\o_l,\s_l}(\xx_{ij}-\yy_{i'j'})&=
<A^{(h_v)}_{\xx(f^-_{ij}),\o_l,\s_l},
B^{(h_v)}_{\xx(f^+_{i'j'}),\o_l,\s_l}>\=\cr
&\=\int\der\yy A^{(h_v)}_{\xx(f^-_{ij})-\yy,\o_l,\s_l}
B^{(h_v)}_{\yy-\xx(f^+_{i'j'}),\o_l,\s_l}\cr}.\Eq(3.51)$$
For instance $A$ and $B$ can be chosen as:
$$\eqalign{& A^{(h_v)}_{\xx,\o_l}= -i\int
d\kk' e^{-i\kk'\xx}
\sqrt{H(a_0^2\sin^2 \hat k_{\s_l})\tilde f_h(k_0,\underline
k'_{\s_l},\hat k_{\s_l}})
{1\over
k_0^2+(2\cos \hat k_{\s_l}\sin \underline k'_{\s_l})^2}\cr
& B^{(h_v)}_{\xx,\o_l,\s_l}=\int d\kk' e^{-i\kk'\xx}
\sqrt{H(a_0^2\sin^2 \hat k_{\s_l})\tilde f_h(k_0,\underline 
k'_{\s_l},\hat k_{\s_l}})
\left[ ik_0+2\o\cos \hat k_{\s_l}\sin \underline 
k'_{\s_l}\right]\cr}\Eq(3.52)$$
and from \equ(3.49)
we easily get \equ(3.94).

By using \equ(3.81) and \equ(3.94) we get, assuming \equ(f)
$$\eqalign{
&\int d\xx_{v_0} |W_{\t,\bP,T,\a}(\xx_{v_0})|\le C^n J_{\t,\bP,\rr,T,\a}
\prod_{v\,\hbox{\ottorm not e.p.}}\cdot\cr
&\cdot C^{\sum_{i=1}^{s_v}|P_{v_i}|-|P_v|-2(s_v-1)}
\g^{{h_v\over 2}\left(\sum_{i=1}^{s_v}|P_{v_i}|-|P_v|-2(s_v-1)\right)}
\;\cdot\cr &\cdot\;
\g^{h_v  \sum_{i=1}^{s_v}\left[ q_\a(P_{v_i}\bs Q_{v_i})+m(P_{v_i}\bs Q_{v_i})
\right] } \g^{-h_v \sum_{l\in T_v}\left[q_\a(f^+_l)+q_\a(f^-_l)+
m(f^+_l)+m(f^-_l)\right]}\Big\},\cr}\Eq(3.100)$$
where
$$\eqalign{
J_{\t,\bP,T,\a}&=\int d\xx_{v_0}
\Big| \Big[ \prod_{i=1}^n d_{j_\a(v^*_i)}^{b_\a(v^*_i)}(\xx_i,\yy_i)
K^{h_i}_{v^*_i}(\xx_{v^*_i})\Big]\cdot\cr
&\cdot \Big\{ \prod_{v\,\hbox{\ottorm not e.p.}} {1\over s_v!}
\Big[\prod_{l\in T_v} \bar\partial^{q_\a(f^-_l)}_{j_\a(f^-_l)}
\bar\partial^{q_\a(f^+_l)}_{j_\a(f^+_l)} 
[d^{b_\a(l)}_{j_\a(l)}(\xx_l,\yy_l)
\hat\dpr^{m_l}g^{(h_v)}_{\o^-_l,\s^-_l;
\o^+_l,\s^+_l}(\xx_l-\yy_l)]\Big]\Big\}\Big|\;.\cr}
\Eq(3.101)$$

In [BM],[BoM] it is proved that
$$\der (\xx_{v_0})=\der\bar\xx\prod_{l\in T}
\der\rr_l\;,\Eq(3.54)$$
where $\rr_l=\xx'_l(t_l)-\yy'_l(s_l)$
and $\xx'_l(t_l), \yy'_l(s_l)$ are interpolated points,
see \equ(f10),\equ(f11),
and $\bar\xx$ is an arbitrary
point of $\xx_{v_0}$. By using \equ(f1), \equ(f2)
we bound dimensionally each
propagator, each derivative and each zero
and we find
$$\eqalign{&J_{\t,\bP,T,\a}\le C^n
\prod_{v\,\hbox{\ottorm not e.p.}} \Big[{1\over s_v!} C^{2(s_v-1)}
\g^{-h_v\sum_{l\in T_v}b_\a(l)+\tilde b_\a(l)}\cdot\cr
&\cdot\g^{-h_v(s_v-1)}\g^{h_v\sum_{l\in
T_v}\left[q_\a(f^+_l)+q_\a(f^-_l)+m(f^+_l)+m(f^-_l)
\right]}\Big]\cr}\; .\Eq(3.57)$$
We find then
$$\int d\xx_{v_0} |W_{\t,\bP,T,\a}(\xx_{v_0})|\le $$
$$C^n L^2\b
|\l|^n
\g^{-h D(P_{v_0})}
\prod_{v\,\hbox{\ottorm not e.p.}} \left\{ {1\over s_v!}
C^{\sum_{i=1}^{s_v}|P_{v_i}|-|P_v|}
\g^{-[-2+{|P_v|\over 2}+z(P_v)]}\right\}\Eq(3.105)$$
where $D(P_{v_0})=-2+m$. The sum over 
$t,\bP,T,\a$ is standard and we refer to [BM], \S 3.15;
at the end the following theorem is proved.
\*
 {\cs Theorem.} {\it Let $h> h_\b\ge 0$. 
If \equ(f) holds
then there exists a constant $c_0$ such that
$$\eqalign{
\sum_{\t\in \TT_{h,n}} &\sum_{\bP\atop |P_{v_0}|=2m} 
\sum_{T\in {\bf T}} \sum_{\a\in A_T}
\int d\xx_{v_0} |W_{\t,\bP,T,\a}(\xx_{v_0})|\le\cr
&\le L^2\b \g^{-hD(P_{v_0})} (c_0\l)^n\;,\cr}\Eq(3.89)$$
where
$$D(P_{v_0})=-2+m\;.\Eq(3.90)$$}

\vskip.5cm
\vskip.5 truecm
\section(5, The flow of running coupling functions)
\vskip.5truecm
\*
\sub(5.3as) {\rm Lemma}
{\it It holds
that $\sum_{\h=\pm}\hat W^h_2(\h{\pi\over\b},\pm {\pi\over 2},k_-)=
\sum_{\h=\pm}\hat W^h_2(\h{\pi\over\b},k_+,\pm {\pi\over 2})=0$}
\vskip.5cm
\proof
We can compute $\hat W_2^h(\kk)$ also by a "single scale" integration;
in fact  $\hat W_2^h(\kk)$ is the kernel of the term
$\psi^{+(\le h)}_\kk\psi^{-(\le h)}_\kk$ on $\VV^{',h}$ defined by

$$e^{-\VV^{',h}}=
\int P(d\psi^{[h,0]})e^{-\VV(\psi^{[h,0]}+\psi^{(<h)})}\Eq(bnm)$$ 
where $P(d\psi^{[h,0]})$ is the fermionic integration 
with propagator
$$g^{[h,0]}(k_0,k_+,k_-)={\chi_{h,0}(\kk)\over
-i k_0+2\cos k_+\cos k_-}\Eq(110a)$$
where 
$$\chi_{h,0}(\kk)=[H(a_0^2\sin^2 k_+)+
H(a_0^2\sin^2 k_-)]
C^{-1}_h(\sqrt{k_0^2+4\cos^2 k_+\cos^2 k_-})\Eq(101a)$$
with $C_h^{-1}=\sum_{k=h}^0 f_k$.
We can write
$W^h_2({\pi\over\b},{\pi\over 2},k_-)$ 
as sum over Feynman 
graphs (see for instance [GM]) and
each Feynman diagram can be written as
$${1\over L^2\b}\sum_{\kk_1}...
{1\over L^2\b}\sum_{\kk_n}g^{[h,0]}(\kk_1)...g^{[h,0]}(\kk_m)
g^{[h,0]}(\sum_{i=1}^m\s^1_i \kk_i+\s^1\kk)...
g^{[h,0]}(\sum_{i=1}^m\s^n_i
\kk_i+\s^n\kk)\Eq(cs)$$
where $n+m$ is an odd number, $\s^j,\s^j_i=0,1,-1$, 
$\s^j+\sum_i\s^j_i$
is an odd integer and $\kk=({\pi\over\b},{\pi\over 2},k_2)$. 
In order to write \equ(cs) we consider
a spanning tree $\TT$ formed by propagators 
connecting all the vertices
of the graphs. We will call
the propagators not belonging to $\TT$ {\it loop lines}
and we write
the momenta of the propagators of $\TT$
as a linear combination of the momenta
of the loop propagators and of the external momentum.
We perform the shift
$k_{+,i}\to k'_{+,i}+{\pi\over 2}$,
and the summation domain is not changed by periodicity. 
The loop propagators become 
$$\bar g^{[h,0]}(\kk')={\bar\chi_{h,0}(\kk')\over
-i k_0+2\sin k'_+\cos k_-}\Eq(110ab)$$
with 
$$\bar\chi_{h,0}(\kk')=[H(a_0^2\cos^2 k'_+)+
H(a_0^2\sin^2 k_-)]
C^{-1}_h(\sqrt{k_0^2+4\sin^2 k'_+\cos^2 k_-})\Eq(101ajjk)$$
Of course $\bar g^{[h,0]}(\kk)$ is {\it odd}
in the exchange $k_0,k_+,k_-\to-k_0,-k_+,k_-$.
On the other hand the momenta of the propagators belonging to $\TT$
becomes
$$\sum_{i=1}^m\s^j_i \kk_i+\s^j\kk
=(\sum_{i=1}^m\s^j_i k_{0,i}+\s^j{\pi\over \b},
\sum_{i=1}^m\s^j_i k'_{+,i}+
(\sum_{i=1}^m\s^j_i+\s^j){\pi\over 2},
\sum_{i=1}^m\s^j_i k_{-,i}+
\s^j k_-)\Eq(jknb)$$
with $(\sum_{i=1}^m\s^j_i+\s^j)$ an odd integer; hence
the propagators belonging to $\TT$
have the form
$${\bar\chi_{h,0}(\kk')\over
-i [\sum_i \s_i k_{0,i}+\s {\pi\over\b}]+
2[(-1)^k \sin(\sum_i\s_i k'_{i,+})][\cos (\sum_i \s_i
k_{-,i}+\s k_-)]}\Eq(110ab)$$
and 
$$\bar\chi_{h,0}(\kk')=
[H(a_0^2\cos^2(\sum_i \s_i k'_{+,i})+
H(a_0^2\sin^2(\sum_i \s_i
k_{-,i}+k_- ))]$$
$$[C^{-1}_h(\sqrt{(\sum_i \s_i k_{0,i})^2+4\sin^2(\sum_i\s_i k'_{i,+})
\cos^2 (\sum_i \s_i
k_{-,i}+\s k_-)} )\Eq(pall)$$
Hence by performing the change of variables
$k_0,k'_+,k_-\to-k_0,-k'_+,k_-$ we find
$$\hat W^h_2({\pi\over\b},{\pi\over 2},k_-)=-
\hat W^h_2(-{\pi\over\b},{\pi\over 2},k_-)\Eq(pzs)$$. 
\*
\sub(5.2) {\it Finite temperature flow.} 
The multiscale analysis defined above 
has the effect that the {\it running coupling functions}
$\bar \d_h(k'_{\s,\o}), Z_h(\bar k'_{\s,\o})$ and
$\l_h(\bar k'_{\s_1,\o_1},
\bar k'_{\s_{2,\o_2}},\bar k'_{\s_3,\o_3})$
verify a recursive relation of the form
$$\d_{h-1}(\bar k'_{\s,\o})=
\d_{h}(\bar k'_{\s,\o})+\b^h_\d(\bar k'_{\s,\o})$$
$${Z_{h-1}(\bar k'_{\s,\o})\over Z_h(\bar
k'_{\s,\o})}=1+\b^h_\x(\bar k'_{\s,\o})\Eq(fl)$$
$$\l_{h-1}(\bar k'_{\s_1,\o_1},\bar k'_{\s_2,\o_2},\bar
k'_{\s_3,\o_3})=
\l_{h}(\bar k'_{\s_1,\o_1},\bar k'_{\s_2,\o_2},\bar
k'_{\s_3,\o_3})+$$
$$\b^h_{\l}(\bar k'_{\s_1,\o_1},\bar k'_{\s_2,\o_2},\bar
k'_{\s_3,\o_3})$$

It is quite easy to prove that,
at temperature not too low, indeed \equ(f) hold.
The proof is done by induction
assuming that \equ(f) holds for $h$ and proving that
it holds also for $h-1$, if $h-1\ge h_\b$ and $\b\le
\exp{\bar c|\l|^{-1}}$, where $\bar c$ is a suitable constant.
In fact
iterating for instance the last of \equ(fl) we find
$$\l_{h-1}(\bar k'_{\s_1,\o_1},
\bar k'_{\s_2,\o_2},\bar k'_{\s_3,\o_3})=\l+\sum_{k=h+1}^0\b^h_{\l}(
\bar k'_{\s_1,\o_1},\bar k'_{\s_2,\o_2},\bar k'_{\s_3,\o_3})\Eq(bop)$$
and from \equ(f) and \equ(3.89) we find, for $|\l|\le {c_2^\l\over 2  c_0^3}$
$$|\sup_{\{k'\},\{\s\}}\b_k^\l(\bar k'_{\s_1,\o_1},\bar k'_{\s_2,\o_2},\bar
k'_{\s_3,\o_3})|\le 2 c^\l_2 \l^2$$
if $c_2^\l>0$ is a bound for the norm of the second order
contribution to $\l_h$.
Hence
$$\sup_{\{k\},\{\s\}}|\l_{h-1}(\bar k_{\s_1,\o_1},\bar k_{\s_2,\o_2},\bar k_{\s_3,\o_3},\bar k_{\s_4,\o_4})|\le
[|\l|+|h|  2 c_2^\l \l^2 ]\;.\Eq(5.19)$$
Then $\sup |\l_{h-1}|\le 2|\l|$ if
$\b<e^{1\over 2 c^\l_2 |\l|}$, as $|h|\le |h_\b|$.
The same argument can be repeated for $\d_h$ and ${Z_{h-1}\over Z_h}$, and \equ(f) holds.
\*
\sub(5.3) {\it Flow of the wave function renormalization.}
To complete the proof of the main Theorem one has to check
that indeed the critical index $\h(k_+)$ or $\h(k_-)$
are non identically vanishing, and this is equivalent to show that
there exists a non vanishing function
$a(\bar k'_{\s,\o})>0$ such that
$$e^{-{a(\bar k'_{\s,\o})\over 2}\l^2h}\le Z_h(\bar k'_{\s,\o})\le 
e^{-2a(\bar k'_{\s,\o}) \l^2h}\Eq(okh0)$$
From the fact that
$\b^h_\x(\bar k'_{\s,\o})=\sum_{n=2}^\io 
\b^{h(n)}_\x(\bar k'_{\s,\o})$ with 
$|\b_\x^{h(n)}(\bar k'_{\s,\o})|\le c_0^n|\l|^n$, as a consequence
of \equ(3.90)
and \equ(f), it is sufficient to find an upper and lower
bound for $\b_\x^{h(2)}$. From an explicit
computation one finds
$$2\o\cos\hat k_\s \b_\x^{h(2)}=24\sum_{\o_1,\o_2,\o_3}
\sum_{\s_1,\s_2,\s_3}{\partial\over\partial \underline k_{\s}}[
\int d\kk_1 d\kk_2 d\kk_3 g^{\le h}_{\s_1,\o_1}(\kk_1)\Eq(neo)$$
$$g^{h}_{\s_2,\o_2}(\kk_2)g^{\le h}_{\s_3,\o_3}(\kk_3)
\d(\kk+\kk_3-\kk_1-\kk_2)\l_h(\bar\kk_\s,\bar\kk_{3,\s_3},\bar
\kk_{1,\s_1})\l_h(\bar\kk_{1,\s_1},\bar\kk_{2,\s_2},\bar
\kk_{\s})]|_{\kk=\bar k_{\s,\o}}.$$
where $g^{\le h}_{\s,\o}=\sum_{k=h_\b}^h
g^{k}_{\s,\o}$.
As the dependence from the momenta of $\l_h$
is quite complex, it is convenient to
replace in the above integral $\l_h$ with
$\l$; if the integral so obtained is 
nonvanishing, the correction
will be surely smaller for $T\ge e^{-(\bar c|\l|)^{-1}}$
for a suitable $\bar c$, as 
$\l_h=\l+O(\l^2|\log \b|)$ from \equ(5.19).
We can choose $\s=I$ for definiteness (the analysis
for $\s=II$ is identical), and we can distinguish
two kind of contributions in the sum over $\s_1,\s_2,\s_3$;
one in which all the 
propagators are $g_I$, and 
the other such that there is at least a propagator $g_{II}$.
The estimate of this second contribution is $O(\l^2\g^h)$,
as it can be immediately checked by dimensional
considerations and applying the derivative in \equ(neo)
over the $g_{II}$ propagators (one can always do that).
We can further simplify the expression we have to compute noting that
$$g^h_{I,\o}(\xx-\yy)=
\int d\kk e^{-i\kk\xx}{H(a_0^2\sin^2 k_-)\tilde f^h(k_0,k_+)
\over-i k_0+2\o\sin k_+}
+\bar g^h_{I,\o}(\xx-\yy)\Eq(hjb)$$
with  
$$|\bar g^h_{I,\o}(\xx-\yy)|\le a_0^{-2}{C_N\g^h\over 1+[\g^h|x_0-y_0|
+\g^h|x_+-y_+|+|x_--y_-|]^N}\Eq(neo1)$$ 
\ie similar to \equ(f1) with an extra $a_0^{-2}$.
We can replace in \equ(neo) the propagators
$g^{(h)}_{I,\o}$ with the first addend in the r.h.s.
of \equ(hjb); if such term will be 
given by a nonvanishing constant, the correction
will be surely smaller at least for $a_0$ large enough.
Hence the dominant contribution to \equ(neo) is given by
$$\sum_{\o_1,\o_2,\o_3}\int d k_{-,1} d k_{-,3}
H(a_0^2\sin^2 k_{-,1})
H(a_0^2\sin^2 k_{-,3})
H(a_0^2\sin^2 (-k_{-,1}+k_{-,3}+k_-)) A\Eq(sdcv)$$
with
$$A=\sum_{\o_1,\o_2,\o_3}
\int dk_{0,1} dk_{0,3}\int d k_{+,1} dk_{+,3}
{f^{\le h}(k_{0,1},k_{+,1})\over -i
k_{0,1}+\o_1  2 k_{+,1}}\times$$
$${f^{\le h}(k_{0,3},k_{+,3})\over -i
k_{0,3}+\o_3 2 k_{+,3}}\partial_{k_+}
{f^{h}(-k_{0,1}+k_{0,3}+k_0,-k_{+,1}+k_{+,3}+k_+)\over -i
(-k_{0,1}+k_{0,3}+k_0)+2 \o_2
(-k_{+,1}+k_{+,3}+k_+)}|_{k_+-\o p_F=k_0=0}\Eq(cbhw)$$
with $\o=\o_1+\o_2-\o_3$; it is easy to check that this term
is indeed non vanishing.
Note also that $A$ is the first non trivial contribution to the critical
index $\h$ of the Schwinger function of
a $d=1$ systems of interacting fermions.
\*
\sub(5.3) {\it Schwinger functions.}
We will note repeat here the analysis of the Schwinger functions
at the temperature scale, as one can proceed as in the
$d=1$ to obtain an expansion for the Schwinger
function once that the expansion for the effective potential
is understood; see 
for instance
[GM]. We only remark that
the $A_I$ and $A_{II}$ in \equ(sf1) and \equ(sf2)
are indeed $O(\l^2)$ as a consequence of
$$\int d k_0 d {k'_+} d k_- g^h_I(k_0, k'_++\o p_F,k_-)=0
\qquad \int d k_0 d k_+ 
d {k'_-} g^h_I(k_0, k_+, k'_-+\o p_F)=0 \Eq(njza)$$

\vskip.5cm
\section(123, Conclusions)
\vskip.5cm
\*
\sub(6.1) {\it Marginal Fermi liquids and Luttinger liquids}.
We can compare the behaviour of the half filled 
Hubbard model with cut-off with other models. We have found that
the wave function renormalization has an anomalous flow
up to exponentially small temperatures, ${Z_{h-1}\over Z_h}=1+O(\l^2)$,
see \equ(okh0); in the case of circular fermi surfaces
one finds instead, see [DR], for $|\l|\le\e$
$${Z_{h-1}\over Z_h}=1+O(\e^2\g^{h\over 2})\Eq(1001)$$
which means that $Z_h=1+O(\l^2)$, up to exponentially small
temperatures; the factor $\g^{h/2}$ in the r.h.s.
of \equ(1001) is an improvement with respect to a
power counting bound and is found by using
a volume improvement
based on the geometrical constraints
to which the momenta close to the Fermi surface 
are subjected. An equation similar to \equ(1001)
holds also for any symmetric smooth Fermi surfaces with non 
vanishing curvature; a proof can be obtained by
combining the results of [BGM]
with Appendix 2 of [DR].

The similarity of the equation for $Z_h$ with its analogous
for one dimensional systems may suggest that the behaviour
of the half-filled Hubbard model with cut-off
up to zero temperature is similar to the one 
of a system of spinless interaction fermions in $d=1$ (the
so called {\it Luttinger liquid} behaviour).
However this is {\it false}; in a Luttinger liquid in fact
one has that
$$\l_{h-1}=\l_h+O(\e^2\g^{h\over 2})\Eq(van)$$
a property known as {\it vanishing of Beta function}.
One can easily check that this cancellation
is not present in the half-filled Hubbard model with cut-off; in fact the dominant
second order
contribution to $\l_h(\bar k_{1,I},
\bar k_{-1,I},\bar k_{1,I})$ containing only $\s=I$
internal lines is
$$\int d\kk' f^h(k_0,k_+) 
f^{\ge h}(k_0,k_+){1\over k_0^2+k_+^2} H(a_0^2\sin^2 k_-)$$
$$[H(a_0^2\sin^2 (k_{1,-}+k_{2,-}-k_-)\l_h \l_h-
H(a_0^2\sin^2 (k_{3,-}-k_{2,-}+k_-)\l_h \l_h]\Eq(lm)$$
where the dependence from $k$ of the $\l_h$ has not been
explicitated. It is clear then that even at the second order the flow of
$\l_h$ is quite complex, and we plan to analyze it
in a future work, in order to understand the leading instabilities. 
Replacing $H$ with $1$ and having $\l_h$
not momentum dependent one recovers the $d=1$ situation
in which the beta function is vanishing.
The theory resembles the theory
of $d=1$ Fermi systems in which each particle has
an extra degree of freedom, the component of the momentum
parallel to the flat Fermi surface, playing the role of a "continuous"
spin index; and it is known that in $d=1$ even 
a spin-${1\over 2}$ index can
destroy the Luttinger liquid behaviour.
\*
\sub(6.2) {\it Marginal Fermi liquid behaviour close
to half filling}. A similar analysis can 
be performed in the case of the Hubbard model
with cut-off {\it close} to half-filling ($\mu=-\e$ 
with $\e$ small and positive);
in such a case the Fermi surface is convex and with finite radius
of curvature but still 
resembles a square with non flat sides
and rounded corners. The propagator has the
form ${\chi(\kk)\over -i k_0+2\cos k_+\cos k_- -\e}$
and it is easy to verify that, if 
$\b<C\min[{1\over\r}]$ where
$\r$ is the radius of curvature of the Fermi surface,
the bounds \equ(f1), \equ(f2) for the single scale
propagator $g^h_{\s,\o}$ still holds; the reason is that,
up to temperatures greater than the inverse of the curvature radius,
the bounds are insensitive to the fact that
the sides of the Fermi surface are not perfectly flat.
One can repeat all the analysis
of the preceding sections and it is found that the Schwinger
functions behave like (1.12), (1.12) for $\l$ small
enough and
$\b<C \min[\min[{1\over\r}],e^{(\bar c|\l|)^{-1}}]$;
in other words 
marginal Fermi liquid behaviour is still found 
close to half filling , up to such temperatures.

On the other hand at lower temperatures, for $\min[{1\over\r}]\le\b\le
e^{(\bar c|\l|)^{-1}}$ (of course assuming $\min[{1\over\r}]\le
e^{(\bar c|\l|)^{-1}}$)
one can apply
the results of [BGM] (valid for any convex symmetric and regular Fermi
surface)
so finding $Z=1+C_\r[\l^2+O(\l^3)]$
where $C_\r$ is a constant which is very large 
for small $\e$ (and diverging at half filling $\e=0$).
Hence, depending on the values
of the parameters, one can have, in the low temperature
region and before the critical temperature, 
two possibilities: the first is to have
only marginal Fermi liquid behaviour $Z=1+O(\l^2\log\b)$,
and the second is to have
marginal Fermi liquid behaviour
up to temperatures $O(\r^{-1})$ and then 
Fermi liquid behaviour up to the critical temperature.

\vskip1cm
{\bf Acknowledgments} This work was partly written during a visit
at the Ecole Polytechnique in Paris. I thank
J. Magnen and V. Rivasseau for their warm ospitality
and useful discussions. I thank also G. Benfatto
and G. Gallavotti for important remarks.
{\baselineskip=12pt
\vskip1cm
\centerline{\titolo References}
\*
\halign{\hbox to 1.2truecm {[#]\hss} &
        \vtop{\advance\hsize by -1.25 truecm \0#}\cr
A&{P.W.Anderson. The theory of superconductivity in high $T_c$ cuprates,
Princeton University Press, Princeton (1997)}\cr
AGD&{A.A.Abrikosov, L.P.Gorkov, I.Dzialoshinsky, {\it Methods
of quantum field theory in statistical physics}, Prentice hall
(1963)}\cr
BGM&{G.Benfatto, A.Giuliani, V.Mastropietro. to appear Annales
Poincare'}\cr
BM& {G. Benfatto, V.Mastropietro. {\it
Renormalization group, hidden symmetries 
and approximate Ward identities in the $XYZ$ model}. 
Rev. Math. Phys. 13 (2001), no. 11, 1323--143}\cr
BoM&{F.Bonetto, V.Mastropietro. {\it 
Beta function and anomaly of the Fermi surface for
a $d=1$ system of interacting fermions in a periodic potential.} 
Comm. Math. Phys. 172 (1995), no. 1,
57--93. }\cr

DAD& {S.Dusuel, F. De Abreu, B. Doucot 
{\it Renormalization Group for 2D fermions with a flat Fermi surface}
cond-mat 0107548}\cr
DR& {M.Disertori, V.Rivasseau. {\it Interacting Fermi liquid in two
dimensions}. Comm. Math. Phys. 215, 251-290 and 291-341 (2000)}\cr
FMRT& {J.Feldman, J. Magnen, V.Rivasseau, E. Trubowitz.
{\it An infinite volume expansion for Many Fermions Green functions}
Helvetica Physica Acta, 65, 679-721 (1992)}\cr
FSW& {A.Ferraz, T.Saikawa, Z.Y. Weng 
{\it Marginal fermi liquid with a two-dimensional 

Patched Fermi surface} cond-mat9908111}\cr
FSL& {J.O. Fjaerested, A. Sudbo, A. Luther. 
{\it Correlation function for a two dimensional electron system with bosonic interactions and a square 
Fermi surface} Phys. Rev. B. 60, 19, 13361-13370 (1999)}\cr
GM& {Gentile,.G, Mastropietro.V. 
{\it
Renormalization group for one-dimensional fermions. 
A review on mathematical results.} Phys. Rep. 352 (2001), no. 4-6,
273--43}\cr
Le& {A. Lesniewski:{\it
Effective action for the Yukawa 2 quantum field Theory}.
{\it Comm. Math. Phys.} {\bf 108}, 437-467 (1987). }\cr
L& {A.Luther,  {\it Interacting electrons on a square Fermi surface}
PRB, vol.30, n.16 (1994) }\cr
M& {D.C.Mattis, {\it Implications 
of infrared instability in a two dimensional electron gas}
PRB, Vol.36, n.1, 1987}\cr
Me& {W.Metzner, {\it Renormalization Groupanalysis of a two dimensional
interacting electron system} Int. Jour. mod. Phys. 16,11, 1889-1898 
(2001)}\cr
R& {V. Rivasseau. {\it 
The two dimensional Hubbard model at half-filling. 
I. Convergent contributions}. 
J. Statist. Phys. 106 (2002), no. 3-4, 693--72}\cr
S&{ Z-X Shen et al, Science 267, 343 (1995)}\cr
Sh&{R. Shankar, Renormalization Group approach to interacting fermions.
Rev. Mod. Phys. 66 (1) 129-192 (1994)}\cr
VZS& {C.M. Varma, Z.Nussinov, W. van Saarloos
{\it Singular Fermi liquids} cond-mat0103393}\cr
VLSAR& {C.M. Varma, P.B. Littlewood, S.Schmitt-Rink, E. Abrahams,
A.E. Ruckestein {\it Phenomenology of the normal state of the Cu-O high $T_c$ superconductors}
Phys. Rev. Lett., 63, 1996, (1989)}\cr
VR& {A.Viroszteck, J.Ruvalds.{\it Nested fermi liquid theory.} Phys Rev. B 42, 4064 (1990) }\cr
ZYD& { A.T.Zheleznyak, V.M. Yakovenko,
I.E. Dzyaloshinskii, {\it Parquet solutions for a flat fermi surface}
Phys. Rev.B, 55, 3200, 1997.}\cr
}
\bye

\vskip.5cm
\section(7,The bibliography)
\vskip1cm
[DjR] D.Djaputra, J.Ruvalds] cond-mat9712053

\bye